\documentclass[aps,pra,twocolumn,superscriptaddress]{revtex4-1}
\usepackage{bm}
\usepackage{graphicx}
\usepackage{amssymb,amsmath,amsbsy,amsgen,amsfonts}
\usepackage{dcolumn}
\usepackage{amsthm}
\usepackage{mathrsfs}
\usepackage{latexsym}
\usepackage{array}
\usepackage{amstext}
\usepackage{epsfig}
\usepackage{epstopdf} 

\usepackage{color}
\usepackage{units}
\usepackage{calrsfs}
\usepackage{verbatim}
\DeclareMathAlphabet\mathbfcal{OMS}{cmsy}{b}{n}

\newcommand{\be}{\begin{equation}}
\newcommand{\ee}{\end{equation}}
\newcommand{\ba}{\begin{array}}
\newcommand{\ea}{\end{array}}
\newcommand{\bqa}{\begin{eqnarray}}
\newcommand{\eqa}{\end{eqnarray}}

\begin{document}


\title{Unidirectional and diffractionless surface plasmon-polaritons \\on three-dimensional nonreciprocal plasmonic platforms}

\author{S. Ali Hassani Gangaraj}
\address{School of Electrical and Computer Engineering, Cornell University, Ithaca, NY 14853, USA}

\author{George W. Hanson}
\address{Department of Electrical Engineering, University of Wisconsin-Milwaukee, 3200 N. Cramer St., Milwaukee, Wisconsin 53211, USA}

\author{ M\'ario G. Silveirinha}
\address{Instituto Superior T\'{e}cnico, University of Lisbon
	and Instituto de Telecomunica\c{c}\~{o}es, Torre Norte, Av. Rovisco
	Pais 1, Lisbon 1049-001, Portugal}

\author{Kunal Shastri}
\address{School of Electrical and Computer Engineering, Cornell University, Ithaca, NY 14853, USA}

\author{Mauro Antezza}
\address{Laboratoire Charles Coulomb (L2C), UMR 5221 CNRS-Universit\'{e} de Montpellier, F-34095 Montpellier, France}
\address{Institut Universitaire de France, 1 rue Descartes, F-75231 Paris Cedex 05, France}

\author{Francesco Monticone}
\email{francesco.monticone@cornell.edu}
\address{School of Electrical and Computer Engineering, Cornell University, Ithaca, NY 14853, USA}

\date{\today}

\begin{abstract}
Light-matter interactions in conventional nanophotonic structures typically lack directionality. 
For example, different from microwave antenna systems, most optical emitters (e.g., excited atoms/molecules, and simple nano-antennas) exhibit quasi-isotropic dipolar radiation patterns with low directivity. Furthermore, surface waves supported by conventional material substrates do not usually have a preferential direction of propagation, and their wavefront tends to spread as it propagates along the surface, unless the surface or the excitation are properly engineered and structured. In this article, we theoretically demonstrate the possibility of realizing \emph{unidirectional and diffractionless surface-plasmon-polariton modes} on a nonreciprocal platform, namely, a gyrotropic magnetized plasma. Based on a rigorous Green function approach, we provide a comprehensive and systematic analysis of all the available physical mechanisms that may bestow the system with directionality, both in the sense of one-way excitation of surface waves, and in the sense of directive diffractionless propagation along the surface. The considered mechanisms include (i) the effect of strong and weak forms of nonreciprocity, (ii) the elliptic-like or hyperbolic-like topology of the modal dispersion surfaces, and (iii) the source polarization state, with the associated possibility of chiral surface-wave excitation governed by angular-momentum matching. We find that three-dimensional gyrotropic plasmonic platforms support a previously-unnoticed wave-propagation regime that exhibit several of these physical mechanisms simultaneously, allowing us to theoretically demonstrate, for the first time, unidirectional surface-plasmon-polariton modes that propagate as a single ultra-narrow diffractionless beam. We also assess the impact of dissipation and nonlocal effects. Our theoretical findings may enable a new generation of plasmonic structures and devices with highly directional response. 

\end{abstract}

\maketitle


\section{Introduction}

At the interface between certain metallic and dielectric materials, light can couple to collective oscillations of the free electrons of the metal, forming a guided wave that is laterally confined to the interface, known as a surface plasmon polariton (SPP) \cite{Maier,Novotny}. Different from conventional guided modes in optical fibers and waveguides, SPP modes are supported by the interface itself, due to a transverse resonance enabled by the opposite optical properties of the interface materials. The peculiar nature of such surface modes, arising from the coupling of electronic and photonic oscillations, enables field localization at scales much smaller than the free-space  wavelength, far beyond what typically achievable with dielectric waveguides, as well as high field enhancement near the interface. 

\begin{figure*}[!hbtp]
	
	\begin{center}
		\noindent \includegraphics[width=5.45in]{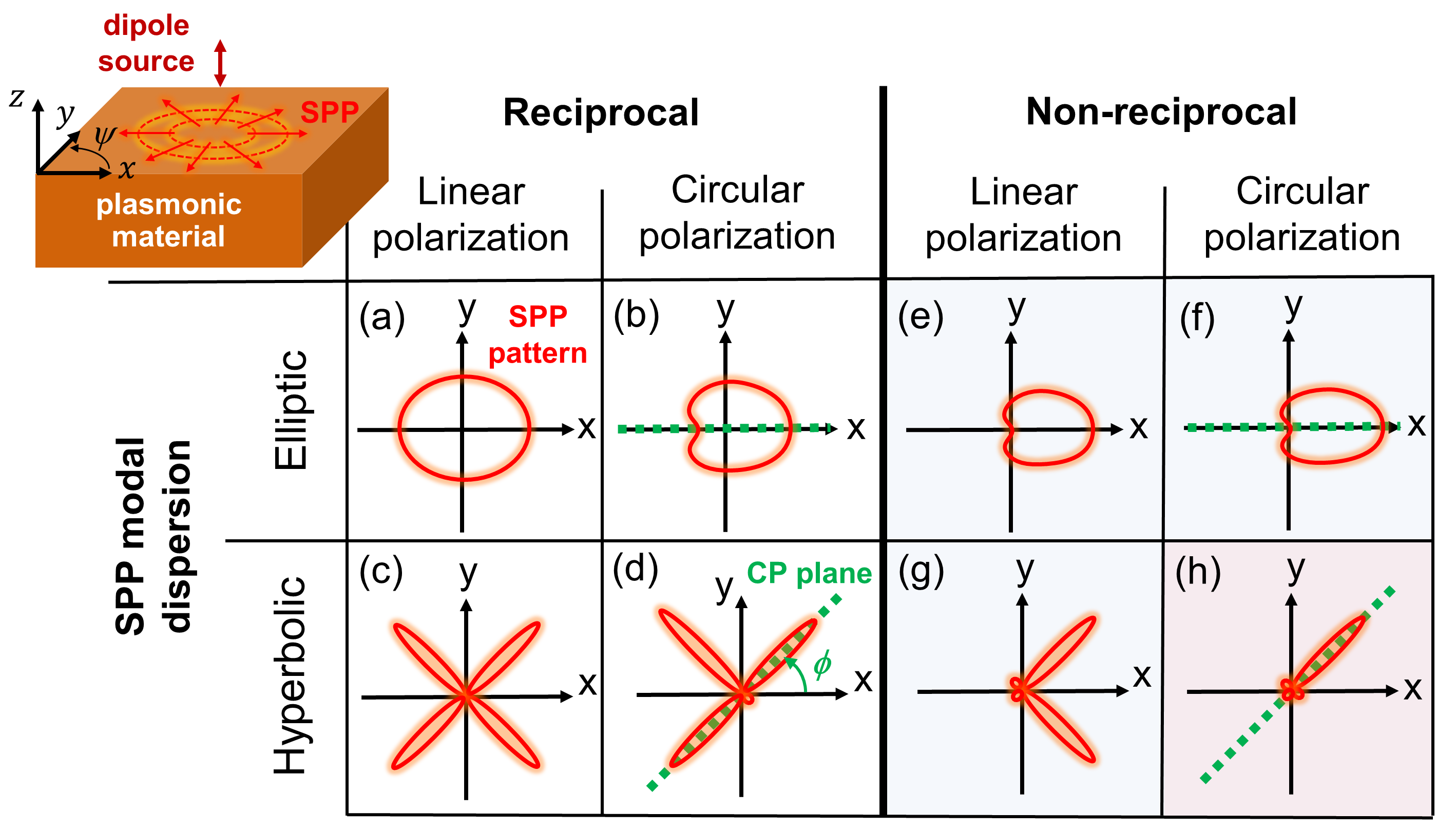}
	\end{center}
	\caption{Different physical mechanisms that influence the directionality of the excitation and propagation of surface plasmon-polaritons (SPPs). Panels (a)-(h) show qualitative sketches of the typical in-plane SPP pattern (surface-wave intensity in different directions on an interface) that may be obtained with a specific combination of source polarization (linear or circular), SPP modal dispersion (elliptic or hyperbolic) and medium properties (reciprocal or strongly nonreciprocal). The inset shows the system configuration under study: a generic plasmonic material occupying the lower half space ($z<0$) forms an interface with a different medium in the upper half space (e.g., free space), where an electric-dipole source is located. The panels corresponding to linear source polarization assume a $z$-oriented dipole, whereas for circularly-polarized emitters the plane of circular polarization is indicated by dashed green lines (the dipole rotates in the plane containing the green line and the $z$-axis). In this work, particularly attention is devoted to panels shaded in blue and red.}
	\label{sketches}
\end{figure*}

Since SPP modes on homogeneous surfaces are slow waves with phase velocity lower than the speed of light in the dielectric environment, they cannot be excited \emph{directly} by an incident propagating plane wave (they can, however, be excited indirectly, by facilitating transverse momentum matching through additional dielectric layers or by suitably structuring the surface with a diffraction grating \cite{Novotny}). 
Conversely, localized emitters and scatterers at near-field distances from the metallic surface can directly launch surface modes. Consider, for example, a nano-emitter with linearly-polarized electric-dipole response, located a short distance above a conventional plasmonic material that is homogeneous, isotropic, and reciprocal (namely, time-reversal symmetry is unbroken). Assuming the linearly-polarized emitter is oriented orthogonal to the interface (inset of Fig. \ref{sketches}), it will excite SPPs that propagate omni-directionally along all in-plane angles, as sketched in Fig. \ref{sketches}(a). This lack of directionality prevents the possibility of launching surface waves along a predetermined direction, and of guiding the SPP energy toward a desired target.

To overcome this issue, in recent years large research efforts have been dedicated to artificial materials and surfaces with extreme anisotropy, with particular attention devoted to so-called hyperbolic structures, which are characterized by effective constitutive-tensor components with opposite signs for orthogonal electric-field polarizations \cite{Kildishev,Gomez-Diaz_3,Hassani_TAP}. In other words, a hyperbolic material or surface may behave as a dielectric or a metal for orthogonal directions of wave propagation. In this scenario, the modes of the system may exhibit hyperbolic dispersion, in contrast to the usual circular/elliptic topology of the dispersion surfaces. By exploiting this property in suitably designed structures, recent works have indeed demonstrated the existence of hyperbolic SPP modes, which propagate on an interface as narrow beams along specific angles determined by the hyperbolic equi-frequency contours (EFCs) of the dispersion surface \cite{Gomez-Diaz_3,Hassani_TAP,Gomez-Diaz_1,Gomez-Diaz_2,Molding_Rev}. However, due to the reciprocal nature of these platforms and their mirror symmetries, such hyperbolic modes lack a preferential left-right and up-down sense of direction, which implies that a linearly-polarized dipole orthogonal to the interface would excite \emph{four} beams propagating along the surface, as sketched in Fig. \ref{sketches}(c). Hence, if point-to-point energy/information transfer is of interest, a reciprocal hyperbolic platform of this type would not be ideal, as surface waves are still guided toward unwanted directions. 

This issue is clearly rooted in the symmetries of the system, namely, time-reversal symmetry (equivalent to reciprocity for dissipationless systems) and mirror symmetries. Indeed, it is evident that, in a reciprocal system, for every forward-propagating mode, there must exist a backward-propagating mode with symmetrical modal distribution and propagation/radiation properties. Therefore, a generic emitter or scatterer is allowed to excite both the forward and the backward modes supported by the reciprocal structure (these modes may be excited with different intensity depending on the specific properties of the emitter/scatterer, but they are both allowed to propagate). To intrinsically forbid the backward mode -- for an arbitrary emitter/scatterer -- it is therefore necessary to break Lorentz reciprocity for wave propagation, which can be done by biasing the system with a physical quantity that is odd under time-reversal, for example, a magnetic field or a linear/angular momentum. The design of advanced nonreciprocal platforms is currently a very active area of research in applied electromagnetics and photonics, with several important practical implications \cite{time_mod}. However, as we discuss in the following, breaking reciprocity is not in itself sufficient for our purposes. Indeed, only strong forms of nonreciprocity enable true unidirectionality, namely, the absence of a backward mode. As an example, the emerging class of artificial materials known as ``photonic topological insulators'' \cite{Haldane,Joannopoulos,Soljacic2014,Ozawa} (the photonic analog of quantum-Hall insulators in condensed-matter physics \cite{Kane}) represent a relevant subclass of strongly-nonreciprocal platforms with unidirectional response. Within this context, in this article we consider another important class of nonreciprocal \emph{continuum} media, i.e., magnetized gyrotropic plasmas, which can be practically implemented using certain natural plasmonic materials, e.g., $n$-doped semiconductors at THz and infrared frequencies, such as $n$-type InSb, under moderate static magnetic bias \cite{Palik, GarciaVidal, SM}. As demonstrated in this article, materials of this type may exhibit both weak and strong forms of nonreciprocity, including topological aspects, accompanied by elliptic or hyperbolic modal dispersion. Thus, we can explore the effect of several physical mechanisms on the propagation of directional SPP modes, as depicted in Figs. \ref{sketches}(e)-(f), based on a naturally-available material platform, without the need to engineer complex photonic crystals or metamaterials.

Another drastically different strategy to select which surface modes get excited on an interface is to suitably design the polarization state of the emitter, such that it matches the properties of the surface modes only in the desired directions. Indeed, as further discussed in the following sections, the lateral confinement of a surface mode directly implies that the mode possesses a transverse component of spin angular momentum, whose sign only depends on the propagation direction, namely, on the sign of the linear momentum \cite{QSH, Lax, Nori, Jacob}. Thanks to this property, known as \emph{spin-momentum locking}, a circularly-polarized emitter would strongly excite only the SPP modes propagating in directions for which their transverse spin matches the spin of the excitation, leading to more directive SPP patterns on the interface, as sketched in Figs. \ref{sketches}(b,d,f,h). In other words, this behavior corresponds to a form of \emph{chiral} asymmetric excitation of surface modes, which has been recently exploited to realize spin-dependent unidirectional emission, scattering, and absorption in reciprocal platforms with transverse light confinement (plasmonic and dielectric waveguides, and nano-optical fibers) \cite{Zoller_Chiral,Lodahl,OC_1,Petersen,OC_2,Mitsch,Feber,Lodhal_2}. 

This type of chiral response is different from spin-dependent effects in (meta)materials systems with chiral constitutive parameters (i.e., magneto-electric coupling) \cite{Lindell}. Indeed, surface modes exhibiting spin-momentum locking do not require chiral material properties, and can be supported by conventional isotropic materials (e.g., a simple plasmonic substrate). It is, in fact, the presence of the circularly-polarized emitter that breaks the mirror symmetry of the system and enables spin-dependent unidirectional effects; conversely, a linearly-polarized emitter would launch surface waves bi-directionally along the surface. It is also clear that the phenomenon of chiral surface-wave excitation is fundamentally distinct from nonreciprocal surface-wave excitation effects, as the latter implies that the backward mode actually does not exist, whereas the former only means that backward and forward modes can be selectively excited due to their opposite angular momentum. This distinction is particularly important for discontinuity problems, where only in the latter case can no back-reflection occur. As seen in the following, these two distinct mechanisms may also be combined in suitable structures, offering additional degrees of freedom to control and tailor the emitter-SPP interaction.



In this paper, we investigate all the physical effects introduced above based on an exact theoretical formulation applied to the relevant case of a nonreciprocal plasmonic substrate illuminated by a generic dipolar emitter. 
We focus on a previously-unnoticed regime of wave propagation supported by a three-dimensional magnetized plasma, which enables the realization of unidirectional \emph{and} diffractionless surface plasmon-polaritons. 
Our investigations reveal an unprecedented degree of control over the excitation and guiding of SPPs, not achievable without considering all the degrees of freedom offered by hyperbolic dispersion, chiral excitation effects, and nonreciprocity.

\section{Overview of theory}

In this section, we provide a brief overview of our theoretical approach to study the interaction between an electromagnetic emitter with arbitrary polarization state and a generic gyrotropic medium. No restrictive assumptions are made on the properties of this medium, which can be dissipative (lossy) and dispersive. The equations governing the electrodynamics of the system can be written in compact form as
\begin{equation}\label{ME}
\textbf{N} \cdot \textbf{f} - i \frac{\partial \textbf{g}}{\partial t} = i \textbf{J}, 
\end{equation}
where the six-vector $ \textbf{f} = [\textbf{E}~ \textbf{H}]^{\mathrm{T}} $ contains the electric and magnetic fields, $ \textbf{g} =  [\textbf{D}~ \textbf{B}]^{\mathrm{T}} $ the electric displacement and magnetic induction fields, and $ \textbf{J} =  [\textbf{j}_e~ \textbf{j}_m]^{\mathrm{T}}  $ the electric and magnetic current densities. The vector fields $ \textbf{f} $ and $ \textbf{g} $ are related by constitutive relations, which may be expressed, in the frequency domain, in the form of a material matrix $ \textbf{M} $. Throughout the paper we assume and suppress a time-harmonic dependence $ e^{-i\omega t} $ for all the fields. For a generic non-magnetic anisotropic medium, we have
\begin{equation}
\textbf{g} = \textbf{M} \cdot \textbf{f},~~ \textbf{M}=\left( {%
	\begin{array}{ccccccccccccccccccc}
	{\boldsymbol{\varepsilon }\left({\mathbf{r},\omega }\right) } & 0 \\
	0 & { {\mu _{0}} \boldsymbol{\rm{I}}}
	\end{array}%
}\right). 
\end{equation} 
The matrix $ \textbf{N} $ in (\ref{ME}) is a linear operator containing the spatial derivatives appearing in Maxwell's equations,
\begin{eqnarray}
& {\boldsymbol{\mathrm{N}}}=\left(
\begin{array}{cc}
\boldsymbol{0} & i\nabla \times \boldsymbol{\mathrm{I}}_{3\times 3} \\
-i\nabla \times \boldsymbol{\mathrm{I}}_{3\times 3} & \boldsymbol{0}%
\end{array}%
\right).
\end{eqnarray}

The frequency-domain dyadic Green function of the system (spatial impulse response of the system) is given by the solution of Eq. (\ref{ME}) for an ideal electro-magnetic point source,
\begin{equation}\label{Green}
\left(\textbf{N} - \omega \textbf{M}\right) \cdot \textbf{G} = i \textbf{I}_{6 \times 6}  \delta(\textbf{r} - \textbf{r}_0 ),
\end{equation}
where $\boldsymbol{\mathrm{r}}$ is the observation point, $\boldsymbol{%
	\mathrm{r}}_{0}$ is the source point, and
\begin{equation} \label{Greendecomp}
\mathbf{G}=\left(
\begin{array}{cc}
\mathbf{G}_{\mathrm{EE}} & \mathbf{G}_{\mathrm{EH}} \\
\mathbf{G}_{\mathrm{HE}} & \mathbf{G}_{\mathrm{HH}}%
\end{array}%
\right)
\end{equation}
is a $6 \times 6$ dyadic (or second-rank tensor) with $ 3 \times 3 $ components $ \textbf{G}_{\alpha, \beta}$, $ \alpha,\beta = \mathrm{E,H}$. 


We are interested in studying the problem of surface-wave excitation by a localized emitter above a substrate or stratified medium (as in the inset of Fig. \ref{sketches}). Assuming an electric point dipole is located in an homogeneous half-space, $ z > 0 $, above a generic planar structure, the electromagnetic field in this region is the superposition of the incident field radiated by the source (primary field), and the field scattered by the substrate (secondary field). The electric Green function associated with the primary field is given by $ (-i\omega\epsilon_0)  \textbf{G}^{\mathrm{inc}}_{\mathrm{EE}}  = \left( {\nabla \nabla  + k_0^2{\bf{\rm{I}}}} \right){\Phi_0}$, where ${\Phi_0} = {{{e^{i{k_0}r}}} \mathord{\left/ {\vphantom {{{e^{i{k_0}r}}} {4\pi r}}} \right. \kern-\nulldelimiterspace} {4\pi r}}$. For a classical dipole with electric dipole moment $\boldsymbol{\gamma }$, the scattered electric Green function, $ \boldsymbol{\mathrm{G}}^s_{\rm{EE}} $ is related to the scattered electric field by $\mathrm{\mathbf{E}}^s =-i\omega \boldsymbol{\mathrm{G}}^s_{\rm{EE}} \cdot {\boldsymbol{\gamma }}$, which can also be expressed in the form of a plane-wave expansion corresponding to the following spatial Fourier integral (Sommerfeld integral) \cite{Novotny}
\begin{eqnarray}\label{GEE}
&\boldsymbol{\mathrm{E}}^{\mathrm{s}} = \iint d
{k}_{x}d {k}_{y} \, \frac{e^{-p_0(z+z')}}{ (2\pi )^{2} 2p_0}\, e^{i\mathbf{k}_{\Vert }\cdot
	(\mathbf{r}-\mathbf{r}')} \, \mathbf{C}\left( \omega
,\mathbf{k}_{\Vert }\right) \cdot \frac{\boldsymbol{\gamma}}{\epsilon_0},
\end{eqnarray}
where $\boldsymbol{\mathrm{k}}_{\parallel
}={k}_{x}\mathbf{\hat{x}}+{k}_{y}\mathbf{\hat{y}}$ is the in-plane wavenumber, ${p_0} =
\sqrt {k_{\parallel}^2 - {k_0^2}}$, $k_0 = \omega/c$, and
\begin{align}\label{CintAp}
 & \mathbf{C}\left( \omega ,\mathbf{k}_{\Vert }\right) = \nonumber \\&  \left( \mathbf{I}_{\Vert }+\widehat{\mathbf{z}}\frac{i\mathbf{k}_{\Vert }%
}{p_0}\right) \cdot \boldsymbol{\mathrm{R}}\left( \omega ,\mathbf{k}_{\Vert }\right) \cdot \left( ip_0\mathbf{k}_{\Vert }\widehat{%
	\mathbf{z}}+k_{0}^{2}\mathbf{I}_{\Vert }-\mathbf{k}_{\Vert }\mathbf{k}%
_{\Vert }\right) 
\end{align}
with $\mathbf{I}_{\Vert }=\widehat{\mathbf{x}}\widehat{\mathbf{x}}+\widehat{\mathbf{y}}\widehat{\mathbf{y}}$. The matrix $ \boldsymbol{\mathrm{R}}\left( \omega ,\mathbf{k}%
_{\Vert }\right) $ in Eq. (\ref{CintAp}) is the reflection matrix that links the tangential components of the fields reflected by the substrate to the corresponding incident fields (see Appendix \ref{APP_A} for additional details). 

Considering a dipolar emitter, at an arbitrary position and with arbitrary polarization state, Eq. (\ref{GEE}) allows calculating, exactly, the field distribution above a generic non-magnetic anisotropic substrate. In particular, the poles of the integrand of (\ref{GEE}) correspond to the discrete spectrum of the eigenmodes supported by the considered structure, for example the SPP modes on a metallic-dielectric interface.

The theoretical formulation above is rigorous and exact; however, to get more physical insight into this problem, a simpler approximate formulation may be developed by assuming that the main radiation channel of the dipolar source is represented by the excitation of a single guided surface mode. Under this assumption, which is typically valid in the cases of interest, it can be shown that the radiation intensity in a certain direction (power radiated by the dipole per unit of angle, i.e., $ U(\psi) = dP_{rad}/d \psi $,  with $ d \psi $ the angular sector of observation) is given by (see \cite{Cherenkov} and Appendix \ref{APP_D})
\begin{equation} \label{rad_patt}
U(\psi) \approx \frac{\omega^2}{16 \pi} \frac{1}{|\boldsymbol{\nabla}_{\boldsymbol{\mathrm{k}}} \omega(\textbf{k})|} \frac{1}{C(\textbf{k})} | \boldsymbol{\gamma}^* \cdot \textbf{E}_{\textbf{k}} (z_0) |^2,
\end{equation}
where the angle $\psi$ is measured from $ +x $-axis in the $xy$-plane, $\textbf{E}_{\textbf{k}} (z_0)$ is the modal electric field, at the location $z_0$ of the source, $C(\textbf{k}) $ is the curvature of the equifrequency contour $\omega(\boldsymbol{\mathrm{k}})=\omega_* $ of the modal dispersion function, at a given frequency $\omega_*$ (e.g., for a circular contour with radius $ |\textbf{k}| $, we have $ C(\textbf{k}) = 1/|\textbf{k}| $). Eq. (\ref{rad_patt}) gives the approximate in-plane radiation pattern of the dipole, corresponding to the in-plane SPP patterns sketched in Fig. \ref{sketches}. This equation reveals that the SPP pattern can be controlled in two ways: (i) By engineering the dispersion function of the relevant surface mode, namely, by controlling (a) the angular dependence of the group velocity, $ | \boldsymbol{\nabla}_{\boldsymbol{\mathrm{k}}} \omega(\textbf{k})| $, and/or (b) the local curvature $ C(\textbf{k}) $ of the equifrequency contour. As mentioned in the Introduction, this can be done by playing with the anisotropy and nonreciprocity of the wave-guiding structure (for example, a hyperbolic dispersion curve exhibits flat asymptotic regions with $ C(\textbf{k}) \approx 0 $ that lead to very directive radiation patterns). (ii) By tailoring the polarization of the dipolar source, which controls the coupling factor $ | \boldsymbol{\gamma}^* \cdot \textbf{E}_{\textbf{k}} (z_0) |^2 $. If a structure is isotropic (and therefore reciprocal), only this latter option is available to control the SPP pattern.

In the following, we use these theoretical formulations to investigate how generic dipolar emitters interact with a nonreciprocal plasmonic substrate. Most importantly, we thoroughly study how (i) the topology of the modal dispersion surface, and (ii) the emitter's polarization state provide the necessary degrees of freedom to control the excitation and guidance of unidirectional and diffractionless SPPs. 

\section{Gyrotropic magnetized plasma as a model nonreciprocal system}

The electromagnetic system under consideration is composed of a homogeneous nonreciprocal material half-space occupying the region $z<0$ covered by an isotropic material in the region $z>0$, where an emitter is located, as in the inset of Fig. \ref{sketches}. 

As a relevant example of a homogeneous nonreciprocal substrate, we consider a gyrotropic material with non-symmetric permittivity tensor $ {\boldsymbol{\epsilon}} = \epsilon_0 ( \epsilon_t {\boldsymbol{\mathrm{I}}}_t  + \epsilon_a \hat{y} \hat{y} + i \epsilon_g \hat{y} \times \boldsymbol{\mathrm{I}}) $, where $ {\boldsymbol{\mathrm{I}}}_t = \boldsymbol{\mathrm{I}} - \hat{y} \hat{y} $, which can be realized as a magnetized plasma with bias magnetic field along the $y$-axis. Interestingly, it has been recently shown that, under certain conditions, continuum gyrotropic materials of this type, with no intrinsic periodicity but with broken time-reversal symmetry, can be understood as examples of topological photonic materials \cite{Silv1sm,Silv2,Si_Re,AliMulti,HeatTransportsm,HM_PRL,HM_AWPL,JPCM}. In the present work, however, we do not focus on the topological properties (Chern invariants, etc.) of magnetized plasmas; instead, we consider this material platform as a model system for studying both strong and weak forms of nonreciprocity, and elliptic or hyperbolic model dispersion.

We assume that the elements of the permittivity tensor of the gyrotropic medium follow the classical dispersion model of a lossy  magnetized free-electron gas \cite{Bittencourt}
\begin{align}\label{bp}
& {\varepsilon _t} = 1 - \frac{{\omega _p^2\left( {1 + i\Gamma /\omega } \right)}}{{{{\left( {\omega  + i\Gamma } \right)}^2} - \omega _c^2}}, ~ {\varepsilon _a} = 1 - \frac{{\omega _p^2}}{{\omega \left( {\omega + i\Gamma } \right)}} \nonumber \\&
 {\varepsilon _g} = \frac{1}{\omega}\frac{{\omega _c^{}\omega _p^2}}{{\omega _c^2 - {{\left( {\omega  + i\Gamma } \right)}^2}}},
\end{align}
where $\omega _{p}$ is the plasma frequency, $\Gamma$ the collision rate associated with damping, $\omega _{c}=-q|B_{0}|/m$ the cyclotron frequency, $q=-e$ the electron charge, $m$ the
effective electron mass, and $B_{0}$ the static magnetic bias. The cyclotron frequency is either positive or negative depending on whether $B_{0}$ is oriented along the $+y$ or $-y$ direction, respectively. As an example, certain $n$-doped semiconductors, such as $n$-type InSb, have a plasma-like response consistent with (\ref{bp}) when subject to a static magnetic bias \cite{Palik, GarciaVidal, SM}. However, we would like to stress that our discussion and considerations in the following sections may qualitatively apply to other nonreciprocal platforms.

\begin{figure}[!tp]
	\begin{center}
		\noindent \includegraphics[width=3.4in]{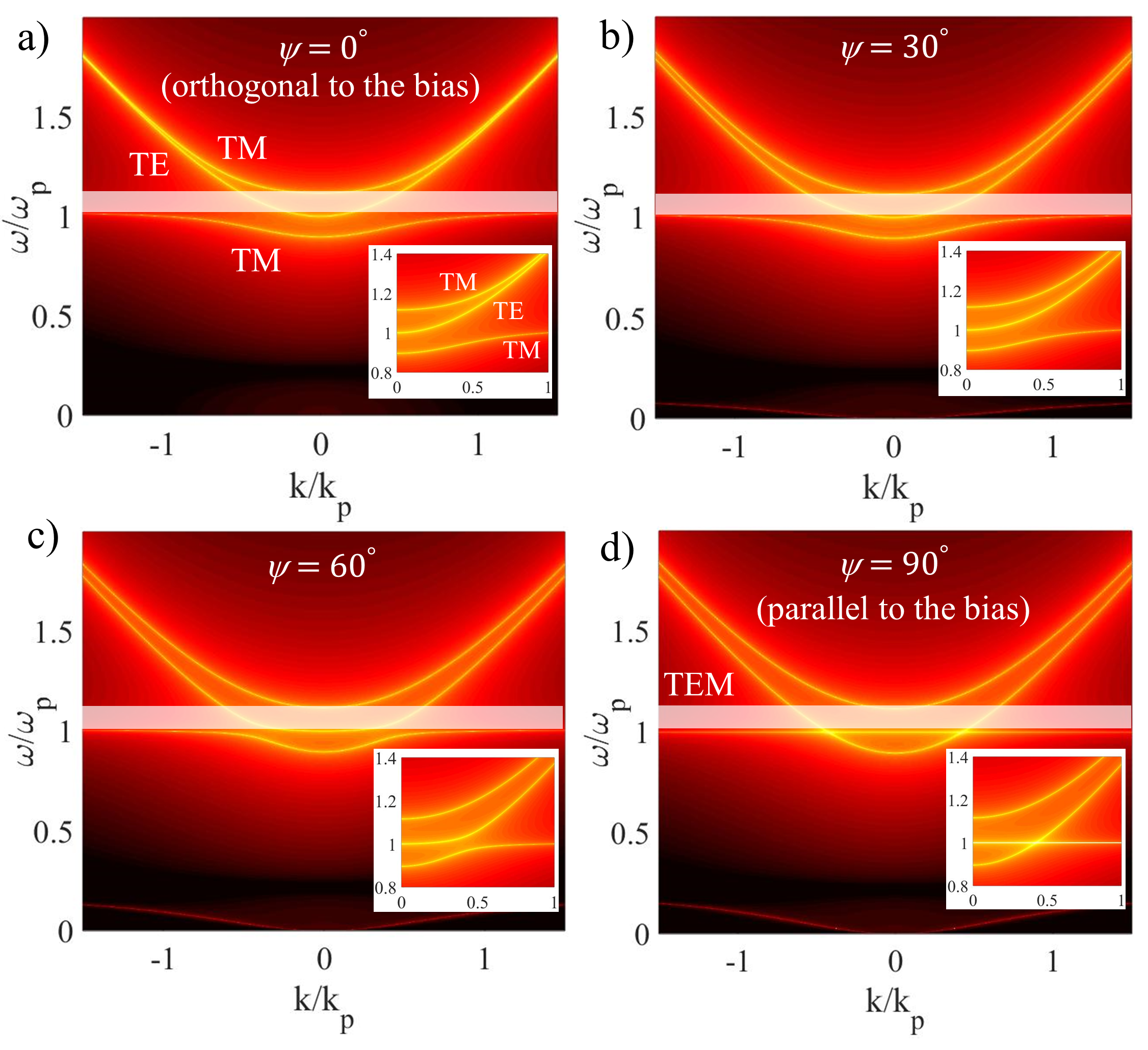}
	\end{center}
	\caption{Band diagram (density plots) for the bulk modes of a magnetized plasmonic material with bias along the $y$-axis, for different propagation directions defined by the angle $\psi$ with respect to the +$x$-axis (see Fig. \ref{sketches}): (a) $ \psi = 0^{\mathrm{o}} $, (b) $ \psi = 30^{\mathrm{o}} $, (c) $ \psi = 60^{\mathrm{o}} $, and (d) $ \psi = 90^{\mathrm{o}} $. The cyclotron frequency is set to $ \omega_c / \omega_p = 0.22 $, where $\omega_p$ is the plasma frequency. $k_p$ is the free-space wavenumber at $\omega_p$. The white horizontal strip indicates the bandgap between TM-like modes. For each panel, the inset provides a zoom around the band gap.}
	\label{bulk_modes}
\end{figure}

A homogeneous three-dimensional magnetized plasma supports several bulk modes of different character (see also Appendix \ref{APP_B}). The band diagram of these bulk modes is shown in Fig. \ref{bulk_modes} for different directions of propagation defined by the angle $\psi$ with respect to the +$x$-axis. For $ \psi = 0^{\mathrm{o}} $ (propagation normal to the bias) there are three bands, as shown in Fig. \ref{bulk_modes}(a). The first and third bands correspond to transverse-magnetic (TM) modes ($ \mathrm{H}_x = 0 $), whereas the second band corresponds to a transverse-electric (TE) mode ($ \mathrm{E}_x = 0 $). The other panels of Fig. \ref{bulk_modes} show how the bulk bands evolve as the angle $ \psi $ is varied. The longitudinal field component of the modes gradually vanish until the modes become TEM for $ \psi = 90^{\mathrm{o}} $ (propagation along the bias). For angles $ \psi > 0^{\mathrm{o}} $, a fourth band appears at low frequencies. 

If we ignore the TE-like modes, we note that the TM-like bands exhibit a common bandgap as the angle is varied, near the plasma frequency $ \omega / \omega_p =1 $, as indicated by the white horizontal strip in Fig. \ref{bulk_modes}. 
The higher and lower frequency limits of this bandgap are given, respectively, by
\begin{equation}
\omega_{H}^2 = \frac{\omega_t^2}{2} \left [  1 + \sqrt{1 - 4 \left(\frac{\omega_p}{\omega_t}\right)^4 }  \right ],
\end{equation}
where $ \omega_t^2 = \omega_c^2 + 2\omega_p^2  $, and
\begin{equation}
\omega_L^2 = \omega_c^2 + \omega_p^2.
\end{equation}

As mentioned in the Introduction, when a plasma-like isotropic and \emph{reciprocal} medium is interfaced with a dielectric medium, TM surface waves are allowed to propagate on the interface, associated with surface-plasmon-polariton modes. Also in the case of a \emph{nonreciprocal} magnetized plasma interfaced with a different medium (interface parallel to the bias axis), TM SPP waves may emerge on the interface, but the dispersion of these modes may be drastically different with respect to the reciprocal case. In the following, we thoroughly study the propagation properties of such SPP modes excited by linearly- and circularly-polarized dipole sources near the surface of a magnetized plasma. We consider different frequency ranges where SPPs can propagate: (i) below the TM bulk-mode bandgap, (ii) within the bandgap, and (iii) above the bandgap. We find that SPP modes exhibit qualitatively different properties in these frequency ranges and, only under specific conditions, unidirectional and diffractionless surface modes can be obtained.

We would also like to note that a more accurate model of a plasmonic material should include the effect of nonlocality (spatial dispersion) for the metal permittivity \cite{nonlocal, Shanhui_arxiv}. However, we show in Supplemental Material \cite{SM} that the results and conclusions of our paper would be essentially unchanged if considering a nonlocal Drude model instead of a local one, as nonlocal effects become important only for very large wavenumbers, which are strongly affected by realistic levels of dissipation. The impact of nonlocality on wave propagation is also strongly dependent on the specific material and configuration under consideration (for example, nonlocal effects would be negligible for gas plasmas \cite{SM}).
	
In addition, we note that the impact of material dissipation and the constraints imposed by passivity should always be considered carefully when studying \emph{terminated} unidirectional channels (e.g., a cavity fed by a one-way input channel) to avoid incorrect predictions and thermodynamic paradoxes \cite{Ishimaru, Mann}.

\section{Linearly-polarized emitter}

We first consider the case of an electric-dipole emitter oscillating linearly in the direction normal to the plasmonic substrate, $ \boldsymbol{ \gamma} = \gamma_z \hat{z} $. In this case, the dipole itself does not break the continuous rotational symmetry around the $z$-axis; hence, if the system was not biased (i.e., reciprocal), the dipole radiation, and the resulting SPP pattern, would necessary be symmetrical in the $xy$-plane [Figs. \ref{sketches}(a),(c)]. The presence of the bias along the $y$-axis clearly breaks this symmetry, determining an increase in the directivity of the dipole emission and SPP pattern [Figs. \ref{sketches}(e),(g)]. The resulting SPP spatial profile, however, largely depends on the allowed angles of propagation of the surface modes, which is determined by the shape of the SPP dispersion function in momentum-space, at a given excitation frequency, as discussed below. 

\subsection{Frequency within and above the bulk-mode bandgap: Asymmetric elliptic-like dispersion}

\begin{figure}[!htbp]
	\begin{center}
		\noindent \includegraphics[width=0.95\columnwidth]{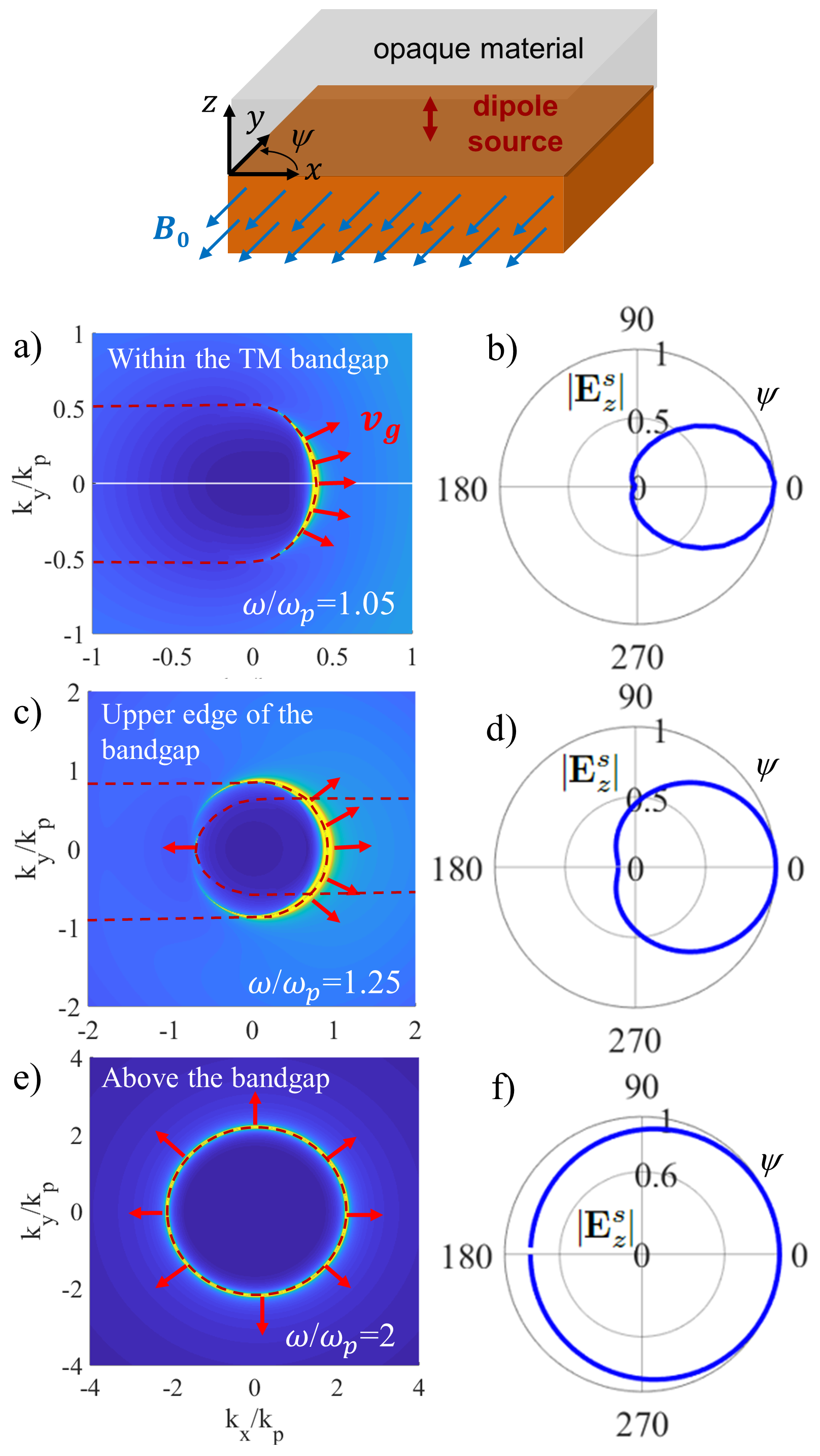}
	\end{center}
	\caption{Surface modes on an interface between a magnetized plasmonic material and an opaque (metallic) isotropic material, as sketched in the inset on top. Left column: Equifrequency contours in $k_x k_y$-space (red dashed lines) for the dispersion function of the TM SPP modes supported by the structure in the inset. The magnetized plasma has cyclotron frequency $ \omega_c / \omega_p = -0.22 $, and the isotropic material has permittivity $ \epsilon_m = -2 $. Three different frequencies $ \omega / \omega_p $ have been considered, within (a) and above (c,e) the bulk-mode bandgap. The colors of the density plots correspond to the magnitude of the integrand in Eq. (\ref{GEE}) (brighter colors mean higher intensity), indicating which portion of the equifrequency contour contributes more strongly to SPP excitation for the chosen source: a dipolar emitter linearly-polarized along the $z$-axis and located a distance $ d = 0.5 c/\omega_p  $ above the surface. The red arrows indicate the main directions of energy flow (i.e., SPP group velocity). Right column: SPP patterns in the $xy$-plane, corresponding to each equifrequency contour on the left, for the same linearly-polarized dipolar emitter. The SPP patterns represent the amplitude of the field $ | \textbf{E}^{s}_z| $, calculated exactly with Eq. (\ref{GEE}), at a fixed radial distance $ 1.2 \lambda_0 $ from the source (where $\lambda_0$ is the free-space wavelength for each panel). In each panel, the fields are normalized to their maximum value.}
	\label{EFS_1}
\end{figure}

When the frequency $\omega$ of the emitter is within the TM bulk-mode bandgap or at higher frequencies ($\omega>\omega_L>\omega_p$), a magnetized plasma layer may support SPP modes on its surface. However, since the surface waves in this configuration are fast waves, with phase velocity larger than the speed of light in vacuum, they tend to lose energy in the form of leaky-wave radiation; therefore, in order to suppress radiation leakage and realize bound surface-wave propagation, we consider an opaque isotropic material above the interface, as sketched in Fig. \ref{EFS_1}. As an example, we consider an interface between a magnetized plasma with $ \omega_c / \omega_p = -0.22 $ (biased in the $-y$ direction) and an isotropic metallic cover with $ \epsilon_m = -2 $ \cite{note2}. The left-column panels of Fig. \ref{EFS_1} show the evolution of the momentum-space equifrequency contours (EFCs) of the dispersion function for the supported SPP mode at different frequencies (red dashed lines). Further details on the dispersion equation of these SPP modes are provided in Appendix \ref{APP_C}.

As seen in Fig. \ref{EFS_1}, the EFC are always more or less \emph{asymmetric} with respect to the in-plane wavevector $k_x$, corresponding to the direction orthogonal to the bias, which is a clear indication of nonreciprocal surface-wave propagation. In a low-loss anisotropic medium/surface, the direction of energy flow is determined by the group velocity, which, different from the isotropic case, does not necessarily coincide with the direction of phase flow determined by the wavevector $ \boldsymbol{\mathrm{k_{\parallel}}}=(\mathrm{k}_x,\mathrm{k}_y) $. Since the group velocity is defined as the gradient of the dispersion function, $ \boldsymbol{v}_g = \boldsymbol{\nabla}_{\boldsymbol{\mathrm{k_{\parallel}}}}\omega(\boldsymbol{\mathrm{k_{\parallel}}}) $, the direction of the group-velocity vector and, therefore, of the SPP energy flow is necessarily orthogonal to the equifrequency contour $\omega(\boldsymbol{\mathrm{k_{\parallel}}})=\omega_* $, at a given frequency $\omega_*$. This direction is indicated in Fig. \ref{EFS_1} by the red arrows, whereas the colors of the density plots indicate which portion of the equifrequency contour contributes more strongly to surface-wave propagation for the considered excitation [the colors correspond to the magnitude of the integrand in Eq. (\ref{GEE})]. The corresponding in-plane SPP patterns around the dipolar source are shown on the right of each EFC panel in Fig. \ref{EFS_1}.

When the frequency of the emitter is within the bulk-mode bandgap, $\omega_L<\omega<\omega_H$, a unidirectional SPP is supported by the material interface, with main direction of propagation toward the positive $x$-axis, whereas zero energy propagates in the opposite direction, as shown in Figs. \ref{EFS_1}(a,b) [similar to the sketch in Fig. \ref{sketches}(e)]. For all frequencies within the bandgap, the EFC is qualitatively similar, yielding unidirectional SPP propagation along the $ +x $-axis with moderate directivity. Figs. \ref{EFS_1}(c,d) show the case of emitter frequency near the upper-edge of the bandgap $\omega\approx\omega_H$ (slightly above it): the EFC of the forward-propagating mode becomes more curved, which determines a broadening of the SPP pattern, and a backward-propagating mode emerges, producing non-zero energy propagation toward the negative $x$-axis (the zero of the SPP profile in this direction transforms into a minimum). As the frequency is further increased, the forward- and backward-mode EFCs tend to become more and more similar and merge into a quasi-symmetric ellipse, as shown in Fig. \ref{EFS_1}(e). As a result, the in-plane SPP pattern in Fig. \ref{EFS_1}(f) is only slightly asymmetric. Finally, for frequencies much higher than the bandgap ($ \omega \gg \omega_H $) the EFS becomes a circle (not shown here) corresponding to isotropic reciprocal SPP propagation. 

These results indicate that the surface of the magnetized plasma supports a \emph{strong form of nonreciprocity} within the bulk-mode bandgap, in the sense that not only is surface-wave propagation asymmetric along the $x$-axis, but the surface modes are also inherently unidirectional, with a zero of the SPP pattern in the $-x$ direction. As mentioned above, the unidirectionality of these SPP modes existing within the bandgap has been recently connected to certain non-trivial topological properties of the biased plasma, which make the SPPs inherently robust to continuous perturbations of the surface, as extensively discussed in \cite{Silv1sm,Silv2,HeatTransportsm,HM_AWPL,HM_PRL,AliMulti,Si_Re}. Conversely, \emph{weak nonreciprocity} is obtained at frequencies above the bandgap, with an SPP pattern that is asymmetric, but not unidirectional. In all the cases studied in this section, however, the ``directivity'' of the SPP beam launched on the surface is low (namely, the width of the main lobe of the SPP pattern is large), which is due to the elliptic-like shape of the EFCs in momentum space. As discussed in the next section, much higher directivity can be achieved at frequencies below the bulk-mode bandgap, where the EFCs of the surface modes are drastically different.

\begin{figure}[!htbp]
	\begin{center}
		\noindent \includegraphics[width=.95\columnwidth]{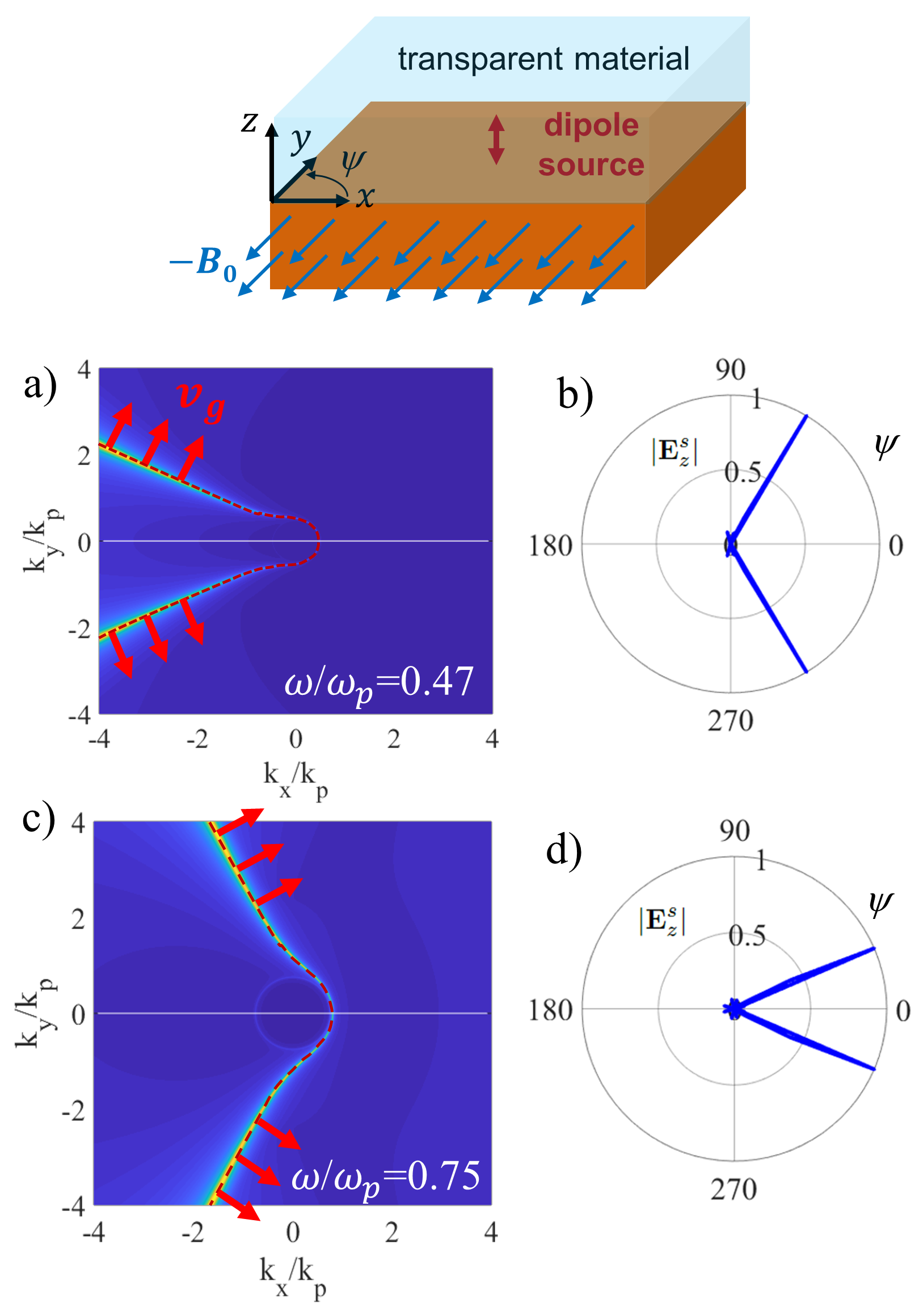}
	\end{center}
	\caption{Surface modes on an interface between a magnetized plasmonic material and a transparent isotropic material (free space), as sketched in the inset on top. The figure is similar to Fig. \ref{EFS_1}, but for two frequencies below the bulk-mode bandgap: (a,b) $ \omega / \omega_p = 0.47  $, and (c,d) $ \omega / \omega_p = 0.75  $. The magnetized plasma has cyclotron frequency $ \omega_c / \omega_p = 0.9 $. The linearly-polarized dipolar emitter is located a distance $ d = 0.05 c/\omega_p  $ above the surface. Left column: Equifrequency contours for the SPP modes (red dashed lines), overlapped to density plots indicating which portion of the equifrequency contour contributes more strongly to SPP excitation (brighter colors) for the chosen source, as in Fig. \ref{EFS_1}. Right columns: Corresponding SPP patterns on the $xy$-plane.}
	\label{EFS_2}
\end{figure}

\subsection{Frequency below the bulk-mode bandgap: Unidirectional semi-hyperbolic dispersion}

When we operate below the bulk-mode bandgap, i.e., $ \omega < \omega_L $, surface modes can still be supported on the interface between the magnetized plasma and an isotropic medium. In this case, we consider again an interface parallel to the $xy$-plane, but the isotropic medium above the plasma is taken to be free space, as sketched in Fig. \ref{EFS_1}. In this case the SPP modes are slow waves, with phase velocity lower than the vacuum speed of light; hence, they do not radiate even though the structure is open.

The EFCs at two different frequencies below the bandgap are shown in Fig. \ref{EFS_2}, left panels. We note that the EFC is drastically different compared to the cases studied in the previous section: the EFC is a single hyperbolic-like contour, strongly asymmetric along the $x$-axis. As in Fig. \ref{EFS_1}, the colors of the density plots indicate how strongly different portions of the equifrequency contour contribute to SPP propagation for the considered excitation. It is therefore evident that the dominant contribution comes from large values of in-plane wave-vector $ \boldsymbol{\mathrm{k_{\parallel}}}$, which corresponds to the asymptotic region of the hyperbolic-like EFC (the intensity of this contribution tails off at much larger values of $ \boldsymbol{\mathrm{k_{\parallel}}}$, as discussed in \cite{SM}). Thus, most of the energy coupled into the SPPs propagates in the same direction determined by the normal to these asymptotes, as indicated by the red arrows in Fig. \ref{EFS_2}. This behavior produces extremely directive surface-wave beams, which propagate -- essentially without diffraction -- only toward the positive $x$-axis, as seen in the SPP patterns in Fig. \ref{EFS_2}, right panels. The angle of propagation of these unidirectional ultra-narrow diffractionless beams can be controlled by varying the excitation frequency, with a wider angle between the beams at lower frequencies [Fig. \ref{EFS_2}(a)].

Interestingly, it can be shown that (see, e.g., \cite{PRA_force,PRB_force,PRB_torque}, for a magnetized plasma interfaced with vacuum, there exists a specific frequency range where these unidirectional semi-hyperbolic SPPs are supported, with upper and lower bounds, $\omega_+$ and $\omega_-$, defined by
\begin{align}  \label{w+_w-}
& \omega_\pm  = \frac{1}{2}
\left(\pm \omega_c + \sqrt{2\omega_p^2 + \omega_c^2} \right).
\end{align}
Furthermore, the dispersion relation of the asymptotic regions of the hyperbolic SPPs (the dominant contribution to the emitter-surface interaction) can be approximated as  $ 2 \omega(\textbf{k}_{\parallel}) = \omega_c \mathrm{cos}(\psi) + \sqrt{ 2\omega_p^2  + \omega_c^2 ( 1 + \mathrm{sin}^2(\psi) ) }$, with $ \psi $ representing the angle between the in-plane SPP wavevector $ \textbf{k}_{\parallel} $ and the $+x $-axis. The frequencies considered for the two examples of unidirectional semi-hyperbolic SPPs in Fig. \ref{EFS_2} indeed lie within the range $ [\omega_-,~ \omega_+] $, and their large-wavenumber behavior is consistent with this approximate dispersion relation. 

We would like to stress that the propagation properties obtained here are drastically different compared to conventional reciprocal hyperbolic surfaces \cite{Gomez-Diaz_3,Hassani_TAP,Gomez-Diaz_1,Gomez-Diaz_2,Molding_Rev}. In the nonreciprocal scenario considered here, we obtained \emph{two} unidirectional ultra-narrow beams propagating along the surface, instead of the usual \emph{four} symmetric beams in the reciprocal case [as sketched in Figs. \ref{sketches}(c),(g)]. Even more interesting would be the ability to excite a \emph{single} ultra-narrow beam; however, this would require breaking the symmetry of the system under parity transformation (mirroring) with respect to the $x$-axis. To achieve this without breaking the transverse invariance of the surface and without introducing chiral material properties (magneto-electric coupling and, more generally, bianisotropy \cite{Lindell}), the only available option is to play with the emitter polarization, consistent with Eq. (\ref{rad_patt}), as discussed below.

\section{Circularly-polarized emitter: Chiral surface-wave excitation}

In this section, we investigate the possibility of realizing unidirectional SPP excitation by suitably engineering the polarization state of the dipolar source, such that also the mirror symmetry of the entire system (source and material structure) is broken, in addition to the broken time-reversal symmetry due to the applied static bias. In particular, by considering emitters that are circularly-polarized on specific planes, we obtain a form of \emph{chiral} SPP excitation that is fundamentally distinct from nonreciprocal forms of excitation; hence, it provides an additional degree of freedom in designing the emitter/SPP interaction. 

\begin{figure}[!htbp]
	\begin{center}
		\noindent \includegraphics[width=.99\columnwidth]{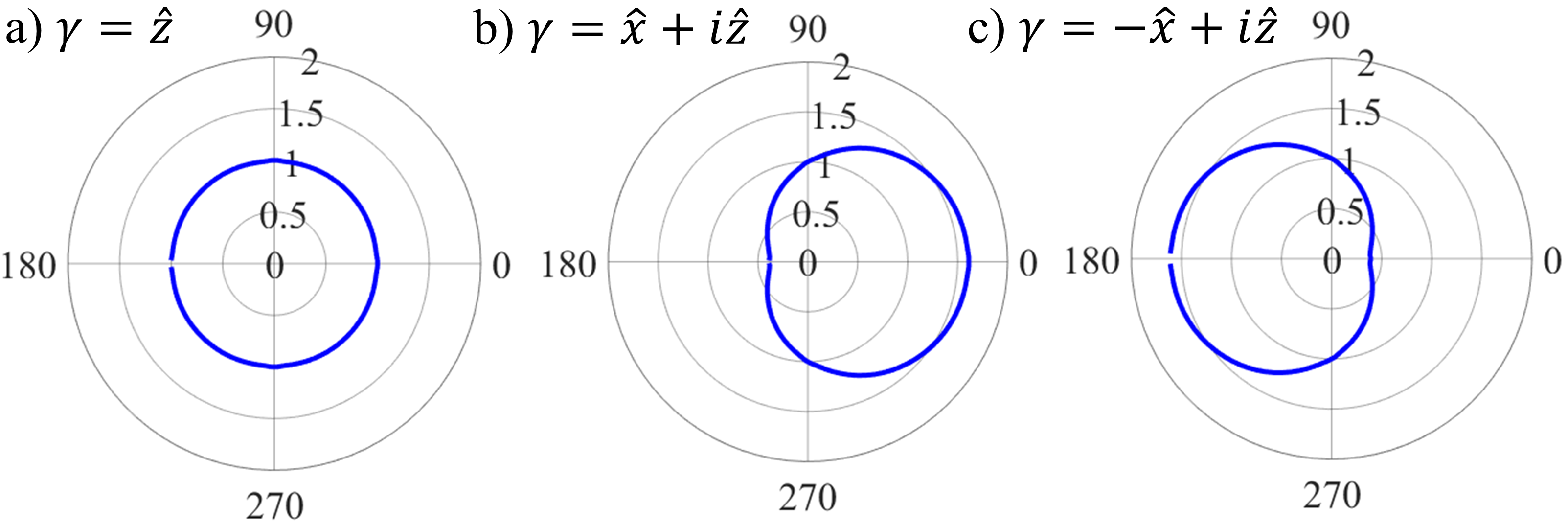}
	\end{center}
 	\caption{SPP patterns in the $xy$-plane on the surface of a non-magnetized plasma, excited by a dipolar emitter with (a) linear polarization $ \boldsymbol{ \gamma} = \hat{z} $, and (b,c) circular polarization, $ \boldsymbol{ \gamma} = \hat{x} + i\hat{z}  $, and $ \boldsymbol{ \gamma} = \hat{x} - i\hat{z}  $, respectively. The SPP patterns represent the amplitude of the field $ | \textbf{E}^{s}_z| $, calculated exactly with Eq. (\ref{GEE}), at a fixed radial distance $ \lambda_0 $ from the source (where $\lambda_0$ is the free-space wavelength). Intensities are normalized to the linear case in (a). The emitter is located a distance $ d = \lambda_0/20 $ above the interface and oscillates at frequency $ \omega/\omega_p = 0.55 $.}
	\label{non_biased}
\end{figure}

\begin{figure}[!htbp]
	\begin{center}
		\noindent \includegraphics[width=2.75in]{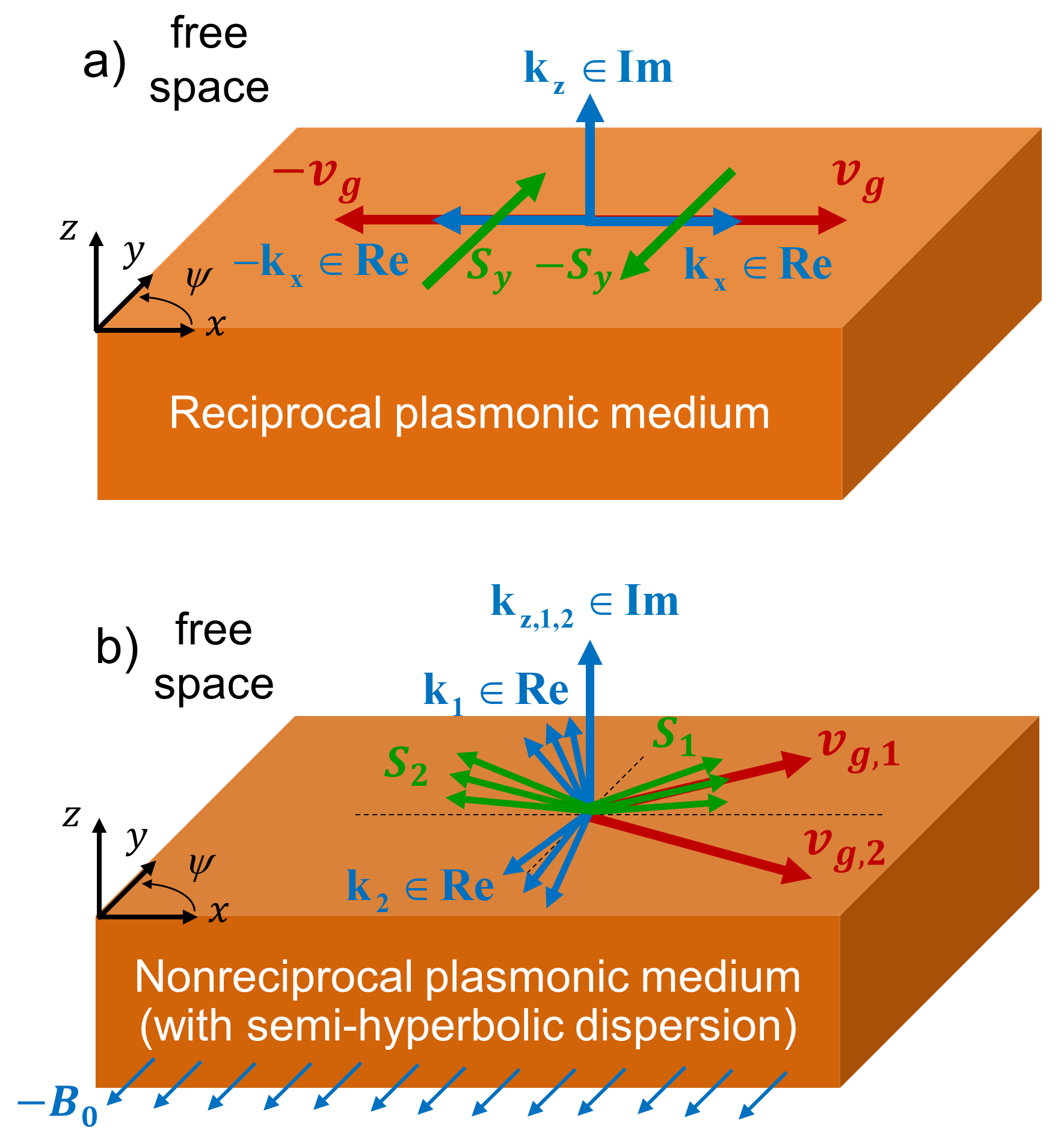}
	\end{center}
	\caption{Schematic of the relevant quantities involved in SPP excitation/propagation: group velocity $\boldsymbol{v}_g$ (red arrows; direction of SPP energy flow), linear momentum $\textbf{k}$ (blue arrows; direction of SPP phase flow if real, and of evanescent decay if imaginary), transverse spin angular momentum $ \textbf{\emph{S}} $ (green arrows; normal to the plane of rotation of the electric field) for a plasmonic surface mode. (a). Two cases have been considered: (a) non-magnetized reciprocal plasma (isotropic bidirectional surface waves), and (b) magnetized nonreciprocal plasma (semi-hyperbolic unidirectional surface waves, i.e., below-the-gap surface modes), interfaced with vacuum. The vectors in panel (b) refer to the dominant large-$k$ asymptotic region of the hyperbolic equifrequency contour in Fig. \ref{EFS_2}.}
	\label{spin}
\end{figure}

To better understand the physical mechanism of this chiral emitter-SPP interaction, consider a simpler case in which the magnetic bias has been turned off, and a dipolar emitter interacts with the plasma-vacuum interface at a frequency $ \omega/\omega_p = 0.55 $. In this scenario, the non-magnetized plasma is simply a reciprocal isotropic material with $ \varepsilon_{t} = \varepsilon_{a} = -2.3,~ \varepsilon_g = 0 $ [given by Eq. (\ref{bp})], which supports reciprocal SPPs when interfaced with a dielectric medium (in this case, vacuum). Let us consider first a dipolar source with linear polarization normal to the interface. In this case there is no preference in the coupling with forward or backward modes; hence, as seen in Fig. \ref{non_biased}(a) [similar to Fig. \ref{sketches}(a)], the SPPs are launched isotropically, propagating along the interface (real in-plane wavevector) and decaying exponentially normal to the interface (imaginary out-of-plane wavevector), as expected. Interestingly, the fact that the wavevector has real and imaginary components along orthogonal directions, as in any evanescent wave, directly implies that the electric field has a longitudinal component in addition to the transverse component (which can be understood from the transversality condition, $\textbf{E} \cdot \textbf{k} = 0,$ applied to an evanescent wave in free space). In addition, the longitudinal and transverse field components have a quadrature phase relation (i.e., $ \pm\pi/2 $ phase difference) with the sign depending on whether the wave is propagating forward or backward in a given direction. The resulting longitudinal rotation of the elliptically-polarized electric field vector, as the wave propagates, indicates that the surface wave carries \emph{transverse} spin angular momentum, as recently recognized in \cite{Nori,QSH,Jacob}. This transverse spin can be written as \cite{QSH},
\begin{equation} \label{spin_momentum}
\textbf{\emph{S}} = \frac{\mathrm{Re}(\textbf{k}) \times \mathrm{Im}(\textbf{k}) }{( \mathrm{Re}(\textbf{k}))^2 },
\end{equation}
which depends only on the direction of propagation of the evanescent wave, and not on the polarization state. As sketched in Fig. \ref{spin}(a), a $+x$-propagating SPP and a $-x$-propagating SPP have equal and opposite transverse spins. Thus, the spin angular momentum of the incident field (i.e., the emitter radiation) can be used to excite, selectively, only the surface waves with transverse spin that matches the spin of the incident field (angular-momentum matching), thereby selecting the direction of the launched surface waves according to Eq. (\ref{spin_momentum}). This behavior is general, not limited to plasmonic interfaces, as any guided surface mode with evanescent tails possesses transverse spin and exhibits spin-momentum-locked propagation \cite{Nori,QSH,Jacob}.

To further understand this behavior from a different viewpoint, consider the source coupling term in Eq. (\ref{rad_patt}), which, for a surface mode in the quasi-static limit, can be approximated as $ | \boldsymbol{\gamma}^* \cdot \textbf{E}_{\textbf{k}} (0) |^2 \approx  \left| \boldsymbol{\gamma}^* \cdot \left(  i\hat{\textbf{k}}_{\parallel} - \hat{\textbf{z}} \right) \right |^2  $ \cite{PRA_force,PRB_force,PRB_torque}, consistent with the aforementioned fact that the electric field is elliptically polarized ($\hat{\textbf{k}}_{\parallel}$ indicates the direction of the in-plane wavevector). If we consider a circularly-polarized emitter,  $ \boldsymbol{ \gamma} \propto \hat{\boldsymbol{ \kappa}} + i \hat{\textbf{z}} $, where $\hat{\boldsymbol{ \kappa}}$ indicates a direction in the $xy$-plane, then $ | \boldsymbol{\gamma}^* \cdot \textbf{E}_{\textbf{k}} (0) |^2 \approx 1 + \mathrm{cos} ( \psi - \phi) $, with $ \hat{\textbf{k}}_{\parallel} \cdot \hat{\boldsymbol{ \kappa}} = \mathrm{cos} ( \psi - \phi) $, where $ \psi  $ is the angle formed by the in-plane modal wavevector and $+x$-axis, and $ \phi $ indicates the polarization plane of the circularly-polarized source (dashed green line in Fig. \ref{sketches}) with respect to the $+x$-axis. Thus, because of spin-momentum locking, the best coupling is always for a mode with in-plane wavevector $ \hat{\textbf{k}}_{\parallel} $ parallel to the in-plane electric-dipole moment, i.e., parallel to $\hat{\boldsymbol{ \kappa}}$. Conversely, the worst coupling (zero/minimum of the SPP pattern) is when $ \hat{\textbf{k}}_{\parallel} $ and $\hat{\boldsymbol{ \kappa}}$ are anti-parallel.

A linearly polarized emitter can be interpreted as the combination of a right-handed circularly polarized (RCP) and a left-handed circularly polarized (LCP) emitter, with equal and opposite values of spin angular momentum (and equal and opposite $\hat{\boldsymbol{ \kappa}}$). According to the discussion above, each sense of rotation gets coupled to either the forward- or backward-propagating mode depending on its spin direction. To further confirm this effect, the SPP patterns produced by circularly-polarized emitters are calculated separately and plotted in Figs. \ref{non_biased}(b),(c) [similar to the sketch in Fig. \ref{sketches}(b)]. It is clear that, depending on the spin of the incident light, either the forward or backward mode is preferentially excited. 
However, from this example it is also evident that reciprocal chiral coupling is not sufficient to realize unidirectional ultra-narrow SPP beams: the SPP pattern exhibits a shallow minimum instead of a zero in the backward direction [compare with Fig. \ref{EFS_1}(a)], and the main SPP beam is broad. This can be overcome by combining chiral excitation and nonreciprocal effects, as discussed in the following.

\subsection{Frequency within the bulk-mode bandgap: Unidirectional elliptic-like surface waves}

\begin{figure}[!htbp]
	\begin{center}
		\noindent \includegraphics[width=2.5in]{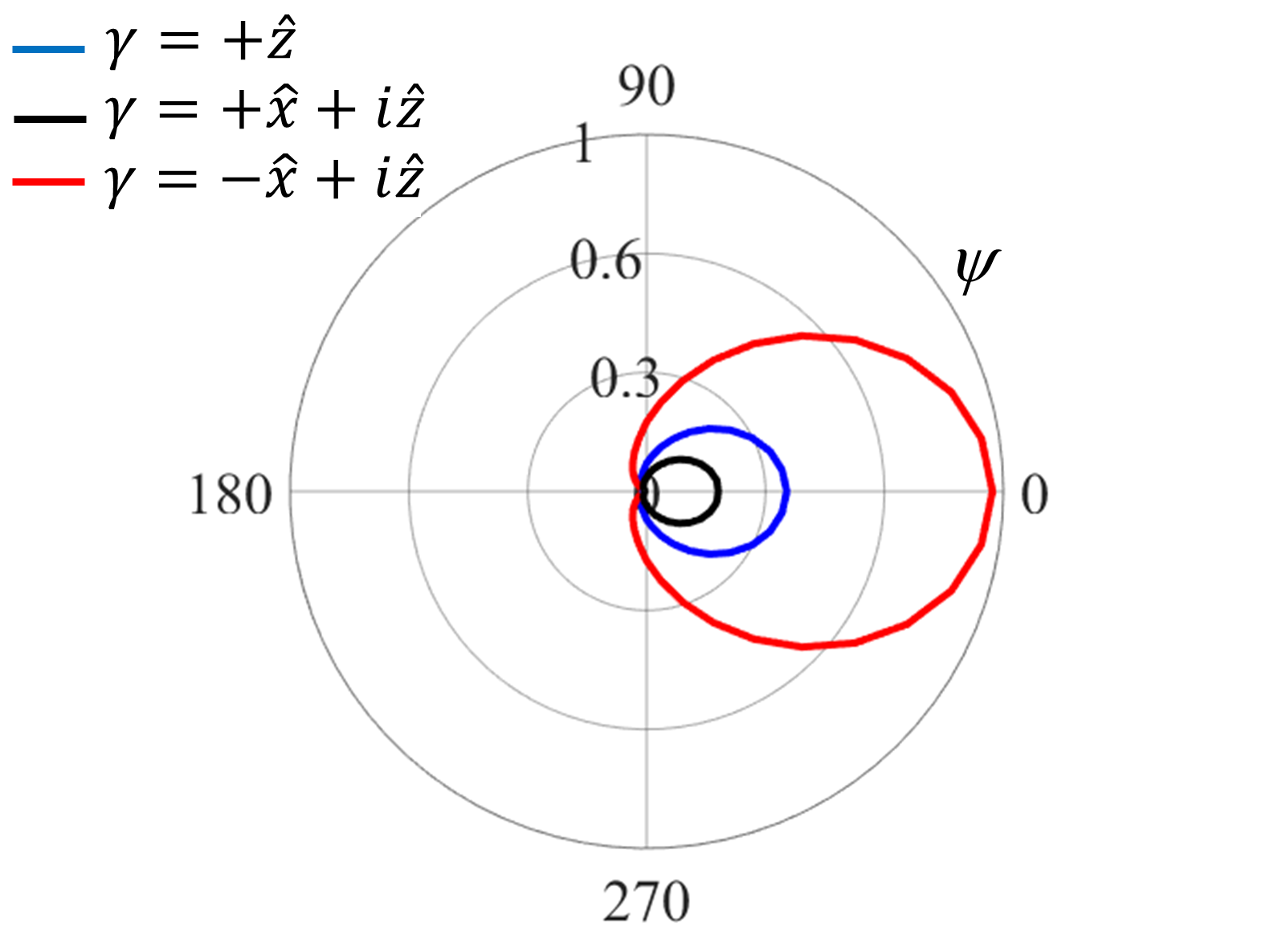}
	\end{center}
	\caption{Elliptic-like SPP patterns in the $xy$-plane on the surface of a magnetized plasma, excited by a dipolar emitter, with linear or circular polarization: $ \boldsymbol{ \gamma} = \hat{z} $ (blue line), $ \boldsymbol{ \gamma} = +\hat{x} + i\hat{z}  $ (black line), and $ \boldsymbol{ \gamma} = -\hat{x} + i\hat{z}  $ (red line). All other parameters are the same as in Fig. \ref{EFS_1}(b).}
	\label{biased_in}
\end{figure}

We now study the effect of the emitter polarization on the excitation of SPPs on a magnetized plasma, when the excitation frequency lies within the bulk-mode bandgap. The parameters of the system are the same as in Fig. \ref{EFS_1}(b), and the excitation frequency is $ \omega/ \omega_p = 1.05 $. Fig. \ref{biased_in} shows the in-plane SPP pattern for different emitter polarization states: linear polarization, $ \boldsymbol{ \gamma} = \hat{z} $ (blue line), and circular polarizations of opposite handedness, $ \boldsymbol{ \gamma} = \hat{x} + i\hat{z}  $ (black line) and $ \boldsymbol{ \gamma} = -\hat{x} + i\hat{z}  $ (red line). For a linearly-polarized dipolar emitter normal to the surface, the SPP profile is the same as in Fig.  \ref{EFS_1}(b): a unidirectional beam with broad angular response, exhibiting a zero in the backward direction, as discussed above. Instead, for a circularly-polarized dipolar emitter with $  \boldsymbol{ \gamma} = \hat{x} + i\hat{z} $ in the $xz$-plane, the incident field would couple more efficiently with a backward-propagating mode due to angular-momentum matching; however, backward propagation is forbidden in this nonreciprocal medium; therefore, the overall energy coupled into the SPP modes is smaller than in the linear case (a weak forward-propagating mode is still excited). Conversely, for a circularly-polarized dipolar emitter with $  \boldsymbol{ \gamma} = - \hat{x} + i\hat{z} $ in the $xz$-plane, the incident field couples more efficiently with a forward-propagating mode, which is the allowed direction of propagation on this nonreciprocal surface. In this case, due to angular-momentum matching between dipolar emitter and forward-propagating SPP, combined with the intrinsic directional preference of the nonreciprocal system, we obtain a much stronger SPP mode launched toward the $+x$-axis, as clearly seen in Fig. \ref{biased_in} (red line) [similar to Fig. \ref{sketches}(f)]. The polarization of the emitter indeed provides an additional degree of freedom to control the excitation of surface modes on a nonreciprocal platform.

\subsection{Frequency below the bulk-mode bandgap: Unidirectional ultra-narrow hyperbolic surface waves}

\begin{figure}[!htbp]
	\begin{center}
		\noindent \includegraphics[width=0.99\columnwidth]{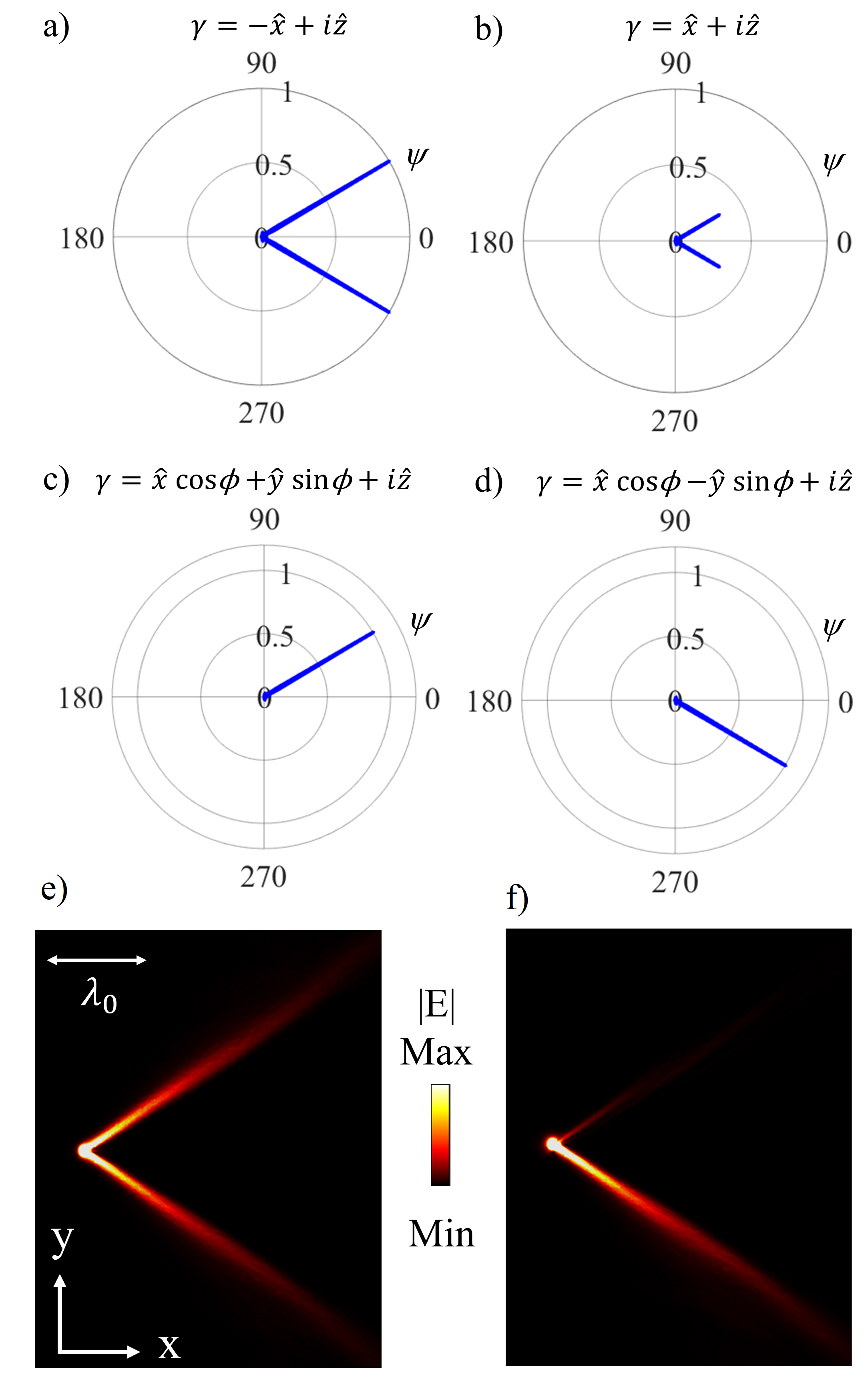}
	\end{center}
	\caption{Semi-hyperbolic SPP patterns in the $xy$-plane on the surface of a magnetized plasmonic material, excited by a dipolar emitter at frequency $ \omega = 0.7 \omega_p $, with different circular-polarization states: (a) $ \boldsymbol{ \gamma} = -\hat{x} + i\hat{z}$, (b) $ \boldsymbol{ \gamma} = \hat{x} + i\hat{z}$, (c) $ \boldsymbol{ \gamma} = \mathrm{cos}(\phi) \hat{x} + \mathrm{sin}(\phi) \hat{y} + i\hat{z}$, and (d) $ \boldsymbol{ \gamma} = \mathrm{cos}(\phi) \hat{x} - \mathrm{sin}(\phi) \hat{y} + i\hat{z}$. The SPP patterns represent the amplitude of the field $ | \textbf{E}^{s}_z| $, calculated exactly with Eq. (\ref{GEE}), at a fixed radial distance $0.7 \lambda_0$ from the source (where $\lambda_0$ is the free-space wavelength). The emitter is located in vacuum, at a distance $ d = 0.05c/\omega_p $ above the magnetized plasma. The plasma cyclotron frequency is $ \omega_c / \omega_p = 0.9 $. The angle considered in panels (c) and (d) is $ \phi = 60^\mathrm{o} $, for this specific example. The intensities of the SPP patterns are all normalized by the same value. Panels (e) and (f) show the electric-field intensity distributions of the launched semi-hyperbolic SPP beams, corresponding to the cases in panels (a) and (d), respectively, obtained via full-wave simulations performed with CST Microwave Studio \cite{CST}. We considered the same parameters as in our exact Green-function calculations, except for the inclusion of moderate dissipation in the plasmonic material, defined by a collision frequency $ \Gamma/\omega_p = 0.003 $. An animation of the simulated time-harmonic electric field, for the case in panel (f), is included as Supplemental Material \cite{SM}.}
	\label{biased_below}
\end{figure}

As discussed in Section IV.B, for an interface between magnetized plasma and air, when the excitation frequency is within the range $[ \omega_- $, $ \omega_+ $], the SPP equifrequency contour has a unidirectional semi-hyperbolic shape, which implies that the interface supports two unidirectional ultra-narrow SPP beams, propagating at a frequency-dependent angle with respect to the $+x$-axis. Figure \ref{spin}(b) depicts the dominant SPP group-velocity vector $\boldsymbol{v}_g$, together with bundles of vectors for linear momentum $\textbf{k}$, and transverse spin angular momentum $\textbf{\emph{S}}$ of the dominant waves. The vector $\textbf{\emph{S}}$ is orthogonal to $\textbf{k}$, which in turn is mostly orthogonal to $\boldsymbol{v}_g$ for a semi-hyperbolic EFC as in Fig. \ref{EFS_2}. Hence, the transverse spin of one of the two excited SPP beams is mostly \emph{parallel} to the group velocity, namely, to the direction of energy flow, whereas the second SPP beam has transverse spin mostly \emph{anti-parallel} to the group velocity, as indicated in Fig. \ref{spin}(b). In this unusual scenario, the effect of emitter polarization is shown in Fig. \ref{biased_below}. Panels (a) and (b) compare the SPP pattern produced by circularly-polarized dipolar emitter of opposite handedness, with $ \boldsymbol{ \gamma} = -\hat{x} + i\hat{z} $ and $ \boldsymbol{ \gamma} = \hat{x} + i\hat{z} $, respectively. In this case, the incident field generated by the emitter with $ \boldsymbol{ \gamma} = - \hat{x} + i\hat{z} $, whose spin has positive $y$-component, couples strongly with the two unidirectional SPP beams, whose transverse spins also have positive $y$-component [see Fig. \ref{spin}(b)]. Conversely, the incident field from the emitter with polarization state $ \boldsymbol{ \gamma} = \hat{x} + i\hat{z} $ couples more weakly to the unidirectional SPP beams, as seen in Fig. \ref{biased_below}(b). Yet, except for a difference in intensity, these SPP patterns look similar to the ones in Fig. \ref{EFS_2} for a linearly-polarized dipole.

Finally, we consider again a dipolar emitter with circular-polarization state, but we now tilt the plane of circular polarization with respect to the $y$-axis. In this way, we are able to completely mismatch (misalign) the spin angular momentum of the incident light with respect to the transverse spin of \emph{only one} of the two beams. As a result, the beam with completely mismatched spin does not get excited by this emitter, while the other beam is launched efficiently. This is possible thanks to the fact that the transverse spins of the two beams are oriented in sufficiently different directions, as sketched in Fig. \ref{spin}(b). As seen in the SPP patterns in Figs. \ref{biased_below}(c) and (d), by playing with the plane of polarization of the dipolar source, defined by the angle $\phi$, we can deliberately select only one of the beams, while the other one is almost completely suppressed [similar to Fig. \ref{sketches}(h)]. By leveraging nonreciprocal effects, hyperbolic dispersion, and angular-momentum matching (chiral coupling), this strategy enables truly unidirectional excitation of surface plasmons, forming a single ultra-narrow beam that propagates -- without diffraction -- on the surface of the structure \cite{note}. Furthermore, the angle of this unidirectional diffractionless beam can be controlled by varying the intensity of the bias or the frequency of the excitation. Particularly striking is the comparison of the delta-function-like SPP patterns in Figs. \ref{biased_below}(c) and (d), with the isotropic or quasi-isotropic SPP patterns obtained with conventional reciprocal plasmonic structures (Fig. \ref{non_biased}).

To further verify these results, we have performed full-wave numerical simulations using a commercial software \cite{CST}. For the cases of vertical linearly-polarized dipole and tilted RCP dipole considered in Figs. \ref{biased_below}(a) and (d), we show in Fig. \ref{biased_below}(e) and (f) the simulated field-intensity distribution, near the source, above a slab of magnetized plasma (biased in the $+y$ direction, and slightly lossy). These results clearly confirm that only one, unidirectional, ultra-narrow, SPP beam is launched on the surface, propagating with little diffraction at an angle dictated by the semi-hyperbolic EFC at the excitation frequency, in striking contrast with the behavior of SPPs on any conventional plasmonic platforms. No energy flows toward the negative $x$-axis due to the inherent unidirectionality of this nonreciprocal platform (see also \cite{SM} for the impact of nonlocality on this unidirectional response). In \cite{SM}, we have also included an animation of the electric-field distribution, corresponding to Fig. \ref{biased_below}(f), which reveals the peculiar rotation of the electric-field vector that produce a component of spin angular momentum along the main direction of energy flow (different from the direction of phase flow), which is responsible for enabling this form of chiral coupling between circularly-polarized emitters and semi-hyperbolic surface waves on a nonreciprocal plasmonic platform.


\section{Conclusion}

\label{SectConcl}

In summary, in this article we have provided a comprehensive theoretical study of surface-plasmon-polariton modes on a nonreciprocal plasmonic platform, namely, a gyrotropic magnetized plasma. Using a rigorous approach based on the exact three-dimensional Green function of the system, we have systematically studied all the available strategies to control the excitation and propagation of unidirectional SPP modes, including (i) the impact of strong and weak forms of nonreciprocity; (ii) the elliptic-like or hyperbolic-like nature of the modal dispersion surfaces, which strongly influences the directivity of the launched SPP wavefront; (iii) the impact of the polarization state of the dipolar source, which may be used to realize a form of chiral excitation of surface waves governed by angular-momentum matching. Most importantly, we have discovered a previously-unnoticed wave-propagation regime supported by homogeneous nonreciprocal plasmas, characterized by two unidirectional semi-hyperbolic propagation channels with distinct spin-polarized propagation properties. This finding allowed us to theoretically demonstrate -- for the first time to the best of the authors' knowledge -- unidirectional \emph{and} diffractionless surface plasmon-polaritons, which propagate as ultra-narrow beams on the two-dimensional surface of a nonreciprocal plasmonic structure.

While our results directly apply to magnetized plasmas and plasmonic materials, the generality of concepts like nonreciprocity, hyperbolic dispersion, transverse spin, and chiral coupling, suggests that the physical insight and general predictions offered by this paper may also qualitatively apply to surface waves supported by other classes of nonreciprocal (meta)materials. We believe that our theoretical findings may open up drastically new opportunities for controlling the excitation and guiding of surface waves, with great practical potential for several applications that benefit from directional wave propagation, including for on-chip point-to-point optical communication and energy transfer, sub-diffraction imaging, and enhanced quantum light-matter interactions.

\section*{Acknowledgments}

F.M. acknowledges support from the National Science Foundation (NSF) with Grant No. 1741694, and the Air Force Office of Scientific Research (AFOSR) with Grant No. FA9550-19-1-0043. M.S. was partially funded by Fundação para a Ciência e a Tecnologia with the grants PTDC/EEITEL/4543/2014 and UID/EEA/50008/2017.


\begin{widetext}
	
	\appendix
	
\section{Reflection matrices and scattered electric field}
\label{APP_A}
As discussed in Section II of the main text, the calculation of the scattered electric field Green function $ \boldsymbol{\mathrm{G}}^s_{\rm{EE}} $ requires determining a reflection matrix that relates the tangential fields reflected by the considered structure to the incident fields. For a gyrotropic material half-space interfaced with an isotropic material, as considered in the main text, by imposing the continuity of the tangential fields at the interface, we can write the reflection matrix in terms of the admittance matrices (further details are provided in \cite{HeatTransportsm,PRA_force}):
\begin{equation} \label{R_matrix}
\boldsymbol{\mathrm{R}}\left( \omega ,\mathbf{k}%
_{\Vert }\right) = \left(\boldsymbol{\mathrm{Y}}_0 + \boldsymbol{%
	\mathrm{Y}}_g\right)^{-1} \cdot \left( \boldsymbol{\mathrm{Y}}_0 -
\boldsymbol{\mathrm{Y}}_g \right),
\end{equation}
where 
\begin{eqnarray}
& \boldsymbol{\mathrm{Y}}_0 = \frac{1}{ i {k}_0 p_0 } \left(
\begin{array}{cc}
-p_0^2 + {k}_x^2 & {k}_x {k}_y \\
{k}_x {k}_y & -p_0^2 + {k}_y^2%
\end{array}%
\right),
\end{eqnarray}
with $p_0^2 = k_x^2 + k_y^2 - k_0^2$, and 
\begin{eqnarray}
\boldsymbol{\mathrm{Y}}_g = & \left(%
\begin{array}{cc}
\frac{\Delta_1 {k}_{t,1}^2 } {{k}_0} & \frac{\Delta_2 {k%
	}_{t,2}^2 } {{k}_0} \\
\frac{\Delta_1 {k}_x {k}_y + i \gamma_{z,1} ( \theta_1 -1 ) {k}_y}{
	{k}_0 } & \frac{\Delta_2 {k}_x {k}_y + i
	\gamma_{z,2} ( \theta_2 -1 ) {k}_y}{ {k}_0 }%
\end{array}%
\right)  \cdot \nonumber\\ & \left( \begin{array}{cc}
{k}_x + i \gamma_{z,1} \Delta_1 & {k}_x + i \gamma_{z,2}
\Delta_2 \\
\theta_1 {k}_y & \theta_2 {k}_y%
\end{array}%
\right)^{-1},
\end{eqnarray}
with 
\begin{eqnarray}
& \Delta_i = \frac{i\varepsilon_g {k}_0^2}{ {k}_0^2 \varepsilon_t -
	( {k}_y^2 + {k}_{t,i}^2 ) }, ~~~ \theta_i =
\frac{ -{k}_{t,i}^2 }{ {k}_0^2 \varepsilon_a - {k}%
	_{t,i}^2 },
\end{eqnarray}
and
\begin{eqnarray}\label{gamma_z}
\gamma^2_{z,i} =& {k}^2_x - \frac{1}{2\varepsilon_t} \left[ \left(
\varepsilon_t \left(  \varepsilon_t + \varepsilon_a  \right)
-\varepsilon_g^2  \right) {k}^2_0 - \left(  \varepsilon_a +
\varepsilon_t  \right) {k}_y^2   \right]  \nonumber\\
& \pm \frac{1}{ 2 \varepsilon_t } \sqrt{  \left[  \left(
	\varepsilon_t \left( \varepsilon_t + \varepsilon_a \right)
	- \varepsilon_g^2 \right) {k}_0^2 - \left( \varepsilon_a + \varepsilon_t \right) {k}_y^2   \right]^2  - 4\varepsilon_t \left[  \left( \varepsilon_t^2 - \varepsilon_g^2 \right) \varepsilon_a {k}_0^4 - 2 \varepsilon_t \varepsilon_a {k}_y^2 {k}_0^2 + \varepsilon_a {k}_y^4 \right]}.
\end{eqnarray}
The admittance matrices $\boldsymbol{\mathrm{Y}}_g$ and $\boldsymbol{\mathrm{Y}}_0$ connect the tangential electric field to the tangential magnetic field for the gyrotropic and isotropic half-spaces, respectively.

We then consider a generic, elliptically polarized, dipolar emitter with electric dipole moment of the form $ \boldsymbol{ \gamma} = \pm \mathrm{cos} (\phi)~ \hat{x} + \mathrm{sin}(\phi) \hat{y}  + i \alpha \hat{z} $, where the angle $\phi$ is measured with respect to the $+x$-axis, indicating the deviation of the polarization plane from the $xz$-plane. For the component of the scattered electric field normal to the material interface, $ \textbf{E}^s_z $, the integrand in Eq. (\ref{GEE}) then becomes
\begin{eqnarray}\label{integrand}
	& \mathbf{ C} \left( \omega ,\mathbf{k}_{\Vert }\right) \cdot\mathbf{ \gamma}|_z = \pm \mathrm{cos}(\phi)  J_{31} ( k_0^2 - k_x^2 ) + J_{32} ( -k_yk_x ) \nonumber \\& + \mathrm{sin}(\phi) \left[ J_{31} (-k_xk_y) + J_{32}( k_0^2-k_y^2 ) \right] + i \alpha \left[J_{31} (ip_0 k_x) + J_{32} ( ip_0 k_y )\right],
\end{eqnarray}
where $ J_{31} = i(  k_xR_{11} + k_y R_{21} )/p_0, ~ J_{32} = i(  k_xR_{12} + k_y R_{22} )/p_0 $ and $ R_{ij},~ i,j=1,2 $ are the elements of the reflection matrix in Eq. (\ref{R_matrix}). For $ \alpha = 1 $, Eq. (\ref{integrand}) gives the scattered field by a circularly-polarized emitter with polarization plane rotated by an angle $\phi$ from the $ xz$-plane. For a linearly-polarized emitter along the $z$-axis, the above equation reduces to $ \mathbf{ C} \left( \omega ,\mathbf{k}_{\Vert }\right) \cdot\mathbf{ \gamma}|_z =  \left[J_{31} (ip_0 k_x) + J_{32} ( ip_0 k_y )\right]$.


\section{Dipole radiation near a wave-guiding structure -- Role of the equifrequency-contour curvature}
\label{APP_D}

The normal $ \hat{\textbf{n}} $ to the equifrequency contour of the relevant mode determines the direction of the group velocity, and hence of the power flow. Consider two closely-spaced points $ \textbf{k}_0 $ and $ \textbf{k}_1 $ on the equifrequency contour separated by a small arc with length $dl$. Let $ \hat{\textbf{n}}_0 $ and $ \hat{\textbf{n}}_1 $ be the corresponding normal vectors, directed along the angular directions $ \psi_0 $ and $ \psi_1
$ with respect to the $+x$-axis in the $xy$-plane. Thus, the power carried by modes with wavevector in the arc with length $dl$ is launched towards a sector with an angular amplitude determined by $ d\psi = \psi_1 - \psi_0 $. From \cite{Cherenkov}, considering for simplicity a single mode, the radiation intensity can be written as
\begin{equation} \label{rad_patt_0}
U(\psi_0) \approx \frac{\omega^2}{16 \pi} \frac{1}{|\boldsymbol{\nabla}_{\boldsymbol{\mathrm{k}}} \omega(\textbf{k})|}  | \boldsymbol{\gamma}^* \cdot \textbf{E}_{\textbf{k}} (z_0) |^2 \frac{dl}{|d\psi|}.
\end{equation}
Using now $ d\psi = \psi_1 - \psi_0 \simeq \mathrm{sin} ( \psi_1 - \psi_0 ) \simeq \hat{\textbf{z}} \cdot ( \hat{\textbf{n}}_0 \times \hat{\textbf{n}}_1 )$, and $ \hat{\textbf{n}}_1 = \hat{\textbf{n}}_0 + \frac{d \hat{\textbf{n}}}{dl}dl $, we obtain $ \frac{ |d\psi| }{dl} = \left |  \hat{\textbf{z}} \cdot \left(  \hat{\textbf{n}}_0 \times \frac{d \hat{\textbf{n}}}{dl}  \right)  \right | $. From the Frenet-Serret formulas (for a curve in the $xy$-plane, i.e., with no torsion) we know that $ \frac{d \hat{\textbf{n}}}{dl} = \pm C  \hat{\textbf{t}}  $, where $C$ is the curvature of the equifrequency contour, and $ \hat{\textbf{t}} $ is the vector tangent to the contour (the sign $ \pm $ depends on the orientation of the curve). From this, we get $ \frac{ |d\psi| }{dl} = \left |  \hat{\textbf{z}} \cdot \left(  \hat{\textbf{n}}_0 \times C \hat{\textbf{t}}  \right)  \right |  = |C| $. Using this result in (\ref{rad_patt_0}), we obtain Eq. (\ref{rad_patt}).

\section{Bulk modes of a three-dimensional nonreciprocal plasmonic medium}
\label{APP_B}
We derive here the exact dispersion equation for the bulk modes of a gyrotropic plasma biased along the $y$-axis (see also,e.g., \cite{HeatTransportsm}). A plane wave in this medium satisfies Maxwell's equation, with $ \nabla \times \boldsymbol{\mathrm{E}} = i\omega \mu_0 \boldsymbol{\mathrm{H}} $ and $ \nabla \times \boldsymbol{\mathrm{H}} = -i\omega \epsilon_0 \boldsymbol{\epsilon} \cdot  \boldsymbol{\mathrm{E}} $, where $ \boldsymbol{\epsilon} $ is the plasma permittivity tensor. The homogeneous wave equation for the electric field in a generic anisotropic material can be written in momentum domain ($ \nabla \rightarrow i\boldsymbol{\mathrm{k}} $) as
\begin{equation}\label{bulk_E}
\boldsymbol{\mathrm{k}} ( \boldsymbol{\mathrm{k}} \cdot \boldsymbol{\mathrm{E}} ) - \mathrm{k}^2 \boldsymbol{\mathrm{E}} + \mathrm{k}_0^2 \boldsymbol{\epsilon} \cdot \boldsymbol{\mathrm{E}} =0,
\end{equation}
where $ \mathrm{k}_0 = \omega / c $ is the free-space wavenumber. We then write the electric field in the form $ \boldsymbol{\mathrm{E}} = \alpha_1 ( \boldsymbol{\mathrm{k}} \times \hat{y} ) + \alpha_2 \boldsymbol{\mathrm{k}}_t + \alpha_3 \hat{y} $, where $ \boldsymbol{\mathrm{k}}_t = \mathrm{k}_x \hat{x} + \mathrm{k}_z \hat{z} $ is the transverse wavenumber with respect to the bias direction. By substituting this expression in Eq. (\ref{bulk_E}), we find that non-trivial (i.e., non-zero) solutions of the homogeneous wave equation should satisfy the dispersion equation:
\begin{equation}
\mathrm{k}_0^4 \left[ \epsilon_a ( \epsilon_t^2 - \epsilon_g^2 )  \right]  - \mathrm{k}_0^2 \left[(-\epsilon_g^2 + \epsilon_t (\epsilon_t + \epsilon_a))\mathrm{k}_t^2 + 2 \epsilon_t \epsilon_a \mathrm{k}_y^2 \right ]  + \left( \mathrm{k}_t^2 + \mathrm{k}_y^2  \right) \left(  \epsilon_t \mathrm{k}_t^2 + \epsilon_a \mathrm{k}_y^2  \right) = 0,
\end{equation}
which implicitly defines the dispersion function, $\omega(\mathbf{k})$, of the bulk modes supported by the magnetized plasma. Furthermore, if we consider bulk-mode propagation along an arbitrary direction, defined by the angle $ \psi $ with respect to $+x$-axis, i.e., $ \mathrm{k}_y = \mathrm{k} ~ \mathrm{sin}(\psi),~ \mathrm{k}_t = \mathrm{k} ~ \mathrm{cos}(\psi)$, the dispersion equation can be re-written as
\begin{equation}
\mathrm{k}_0^4 \left[ \epsilon_a ( \epsilon_t^2 - \epsilon_g^2 )  \right]  - \mathrm{k}_0^2 \left[(-\epsilon_g^2 + \epsilon_t (\epsilon_t + \epsilon_a))\mathrm{cos}(\psi)^2 + 2 \epsilon_t \epsilon_a \mathrm{sin}(\psi)^2 \right ]\mathrm{k}^2  +  \mathrm{k}^4  \left(  \epsilon_t \mathrm{cos}(\psi)^2 + \epsilon_a \mathrm{sin}(\psi)^2  \right) = 0.
\end{equation}
The bulk-mode band diagrams for different angles $ \psi $ are shown in Fig. \ref{bulk_modes} of the main text.


\section{Dispersion equation of the SPP modes}
\label{APP_C}

Consider a magnetized-plasma half-space interfaced with an isotropic-medium half-space at $z = 0$. Due to the translational symmetries of the system, the modal fields in the region $z \gtrless0$ vary as $ e^{ik_xx} $ and $e^{ik_yy}$ along the interface. In the gyrotropic region, the fields can be written as a superposition of two plane waves, modal solutions in the bulk of the gyrotropic medium, with wavevector components $\boldsymbol{%
\mathrm{k}}_i = \boldsymbol{\mathrm{k}}_{t,i} + {k}_y \mathbf{{\hat{y}%
}}$, with $\boldsymbol{\mathrm{k}}_{t,i} = {k}_x \mathbf{\hat{x}} + {k}_{z,i} \mathbf{\hat{z}} $ ($i = 1,2 $). Surface modes decay exponentially away from the interface, so we set ${k}_{z,i} = -i\gamma_{z,i}$ such that $\mathrm{Re}\left( \gamma_{z,i} \right) > 0 $.
For this plane-wave superposition, the electric field can be written in the form
\begin{eqnarray}  \label{E_gyro_bulk}
& \boldsymbol{\mathrm{E}} = \left( \Delta_1
\boldsymbol{\mathrm{k}}_1 \times \mathbf{\hat{y}} + \mathbf{k}_{t,1}
+ \theta_1 {k}_y \mathbf{\hat{y}}
\right) {A}_1 e^{ \gamma_{z,1}z }  + \left( \Delta_2 \boldsymbol{\mathrm{k}}_2 \times
\mathbf{\hat{y}}
+ \mathbf{k}_{t,2} + \theta_2 {k}_y \mathbf{\hat{y}} \right) {A%
}_2 e^{ \gamma_{z,2}z },
\end{eqnarray}
where the variation along $x$ and $y$ is omitted, and ${A}_i~ (i=1,2) $ are expansion coefficients. The corresponding magnetic field can be found using $ \boldsymbol{\mathrm{H}} =  \boldsymbol{\mathrm{k}} \times \boldsymbol{\mathrm{E}} / \omega \mu_0  $. Similarly, we can write a generic field in the isotropic region as follows

\begin{eqnarray}
\boldsymbol{\mathrm{E}} &= - \left[ \mathrm{B}_1 \boldsymbol{\mathrm{k}}_0 \times \hat{z} + \mathrm{B}_2 \boldsymbol{\mathrm{k}}_0 \times ( \boldsymbol{\mathrm{k}}_0 \times \hat{z} )   \right] e^{-p_0 z} 
\omega \mu_0 \boldsymbol{\mathrm{H}}  \nonumber \\& = - \left[   \mathrm{B}_1 \boldsymbol{\mathrm{k}}_0 \times ( \boldsymbol{\mathrm{k}}_0 \times \times \hat{z} ) - \mathrm{B}_2 \frac{\omega^2}{c^2} \epsilon_d ( \boldsymbol{\mathrm{k}}_0 \times \times \hat{z} ) \right] e^{-p_0 z},
\end{eqnarray}
where $ \boldsymbol{\mathrm{k}}_0 = \mathrm{k}_x \hat{x} + \mathrm{k}_y \hat{y} + i p_0 \hat{z} $, $ p_0 = \sqrt{    \mathrm{k}_x^2 + \mathrm{k}_y^2 - \epsilon_d \omega^2 / c^2 } $, and ${B}_i~ (i=1,2) $ are expansion coefficients. By imposing electromagnetic boundary conditions (matching the tangential fields) at the interface, we get the following system of equations (see also,e.g., \cite{HeatTransportsm})

\begin{eqnarray}
& \left(\begin{array}{cccc}
\mathrm{k}_x + i \gamma_{z,1}\Delta_1 & \mathrm{k}_x + i \gamma_{z,2} \Delta & \mathrm{k}_y & \frac{\mathrm{k}_x i p_0 c}{\omega} \\
\theta_1 \mathrm{k}_y & \theta_2 \mathrm{k}_y  & -\mathrm{k}_x & \frac{\mathrm{k}_yip_0 c}{\omega} \\
\Phi_1  & \Phi_2  & \mathrm{k}_x i p_0  & \frac{-\epsilon_d \mathrm{k}_y \omega}{c} \\
-\Delta_1 \mathrm{k}^2_{t,1} &  -\Delta_2 \mathrm{k}^2_{t,2} & \mathrm{k}_y i p_0 & \frac{\epsilon_d \mathrm{k}_x \omega}{c} 
\end{array}\right) \cdot 
\left(\begin{array}{c}
\mathrm{A}_1 \\ \mathrm{A}_2 \\ \mathrm{B}_1 \\ \mathrm{B}_2 \frac{\omega }{c}
\end{array}\right) = \boldsymbol{0}_{4 \times 1},
\end{eqnarray}
where $ \Phi_i = \Delta_i \mathrm{k}_x \mathrm{k}_y + i \gamma_{z,i} (\theta_i -1 )\mathrm{k}_y $, $(i=1,2)$. By setting the determinant equal to zero, one finds the dispersion equation of the SPP modes supported by a planar homogeneous interface between a gyrotropic magnetized plasma and an isotropic medium.


\end{widetext}



\begin{thebibliography}{99}


\bibitem{Maier} S. A. Maier, \emph{Plasmonics: fundamentals and applications}, Berlin: Springer (2007).

\bibitem{Novotny} L. Novotny, B. Hecht, \emph{Principles of nano-optics}, Cambridge: Cambridge University Press (2006).
%
%
%
%





\bibitem{Kildishev} V. P. Drachev, V. A. Podolskiy, and A. V. Kildishev, \emph{Hyperbolic metamaterials: New physics behind a classical problem}, Opt. Exp., vol. 21, no. 12, pp. 15048–15064, (2013).

\bibitem{Gomez-Diaz_3}J. S. Gomez-Diaz, M. Tymchenko, A. Au, \emph{Hyperbolic plasmons and topological transitions over uniaxial metasurfaces}, Phys. Rev. Lett., vol. 114, no. 23, p. 233901, (2015).

\bibitem{Hassani_TAP} S. Ali Hassani Gangaraj, T. Low, A. Nemilentsau, G. W. Hanson, \emph{Directive surface plasmons on tunable two-dimensional hyperbolic metasurfaces and black phosphorus: Green’s function
	and complex plane analysis}. IEEE Trans Antennas Propag 2016;65:1174–86 (2017).

\bibitem{Gomez-Diaz_1} J. S. Gomez-Diaz, M. Tymchenko, and A. Alu, \emph{Hyperbolic plasmons and topological transitions over uniaxial metasurfaces}, Phys. Rev. Lett., vol. 114, no. 23, p. 233901, (2015).

\bibitem{Gomez-Diaz_2} J. S. Gomez-Diaz, A. Alu, \emph{Flatland optics with hyperbolic metasurfaces}, ACS Photon, 3:2211–24 (2016).

\bibitem{Molding_Rev} S. A. Hassani Gangaraj, F. Monticone, \emph{Molding light with metasurfaces: from far-field to near-field interactions}, Nanophotonics vol. 7, issue 6 (2018).

\bibitem{time_mod} D. L. Sounas, A. Alu \emph{Non-reciprocal photonics based on time modulation}, Nature Photonics, vol. 11, pp. 774–783 (2017).


\bibitem{Haldane} S. Raghu and F. D. M. Haldane, \emph{Analogs of quantum-Hall-effect edge states in photonic crystals}, Phys. Rev., vol. 78, Art. no. 033834 (2008).

\bibitem{Joannopoulos} Z. Wang, Y. D. Chong, J. D. Joannopoulos, M. Solja\v{c}i\'{c},
\emph{Reflection-free one-way edge modes in a gyromagnetic photonic crystal}, Phys. Rev. Lett. 100, 013905 (2008).

\bibitem{Soljacic2014} L. Lu, J. D. Joannopoulos, and M. Solja\v{c}i\'{c},  \emph{Topological photonics,} Nat. Photonics 8, 821 – 829 (2014).

\bibitem{Ozawa} T. Ozawa, H. M. Price, A. Amo, N. Goldman, M. Hafezi, L. Lu, M. Rechtsman, D. Schuster, J. Simon, O. Zilberberg, I. Carusotto, \emph{Topological Photonics}, arXiv:1802.04173 (2018).


\bibitem{Kane} M. Z. Hasan and C. L. Kane, \emph{Colloquium : Topological insulators,} Rev. Mod. Phys. 82, 3045–3067 (2010).

\bibitem{Palik} E. Palik, R. Kaplan, R. Gammon, H. Kaplan, R.
Wallis, and J. Quinn, \emph{Coupled surface
	magnetoplasmon-optic-phonon polariton modes on InSb}, Phys. Rev. B
{\bf 13}, 2497, (1976).

\bibitem{GarciaVidal} E. Moncada-Villa, V. Fernandez-Hurtado, F. J. Garcia-Vidal, A.
Garcia-Martin, and J. C. Cuevas, \emph{Magnetic field control of
	near-field radiative heat transfer and the realization of highly
	tunable hyperbolic thermal emitters}, Phys. Rev. B {\bf 92}, 125418,
(2015).

\bibitem{SM} See Supplemental Material for a time-harmonic animation of the electric-field distribution of a unidirectional ultra-narrow SPP beam, corresponding to Fig. \ref{biased_below}(f); an analysis of the impact of material nonlocality and dissipation on unidirectional SPP propagation; realistic parameters for an example of potential physical implementation of the proposed platform; and an analysis of the behavior of the semi-hyperbolic surface modes in momentum space [Fig. \ref{EFS_2}(c)] for large wavenumbers.


\bibitem{Lax} M. Lax, W. H. Louisell, W. B. McKnight, \emph{From Maxwell to paraxial wave optics}, Phys. Rev. A 11, 1365–1370 (1975).

\bibitem{Nori}K. Y. Bliokh, F. Nori, \emph{Transverse spin of a surface polariton}. Phys. Rev. A 85, 061801 (2012).

\bibitem{QSH} K. Y. Bliokh, D. Smirnova, F. Nori, \emph{Quantum spin Hall effect of light}, Science 348, 1448-1451 (2015).

\bibitem{Jacob} T. Van Mechelen and Z. Jacob, \emph{Universal spin-momentum locking of evanescent waves,} Optica 3, 118 (2016).

\bibitem{Zoller_Chiral} H. Pichler, T. Ramos, A. J. Daley, P. Zoller, \emph{Quantum optics of chiral spin networks}, Phys. Rev. A 91, 042116 (2015).

\bibitem{Lodahl} P. Lodahl, S. Mahmoodian, S. Stobbe, A. Rauschenbeutel, P. Schneeweiss, J. U. Volz, H. Pichler, P. Zoller, \emph{Chiral quantum optics}, Nature 541 (7638), 473 - 480 (2017).

\bibitem{OC_1} F. J. Rodriguez-Fortuno, G. Marino, P. Ginzburg, D. O'Connor, A. Martinez, G. A. Wurtz, A. V. Zayats, \emph{Near-Field Interference for the Unidirectional Excitation of Electromagnetic Guided Modes}, Science 340, 328–330 (2013).

\bibitem{Petersen} J. Petersen, J. Volz, A. Rauschenbeutel, \emph{Chiral nanophotonic waveguide interface based on spin-orbit interaction of light}, Science 346, 67 - 71 (2014).

\bibitem{OC_2} D. O'Connor, P. Ginzburg, F. J. Rodríguez-Fortuño, G. A. Wurtz, A. V. Zayats, \emph{Spin–orbit coupling in surface plasmon scattering by nanostructures}, Nat. Commun. 5, 5327 (2014).

\bibitem{Mitsch} R. Mitsch, C. Sayrin, B. Albrecht, P. Schneeweiss, A. Rauschenbeutel, \emph{Quantum state-controlled directional spontaneous emission of photons into a nanophotonic waveguide}, Nat. Commun. 5, 5713 (2014).

\bibitem{Feber} B. le Feber, N. Rotenberg, L. Kuipers, \emph{Nanophotonic control of circular dipole emission}, Nat. Commun. 6, 6695 (2015).

\bibitem{Lodhal_2} I. Sollner, S. Mahmoodian, S. L. Hansen, L. Midolo, A. Javadi, G. Kirsanske, T. Pregnolato, H. El-Ella, E. H. Lee, J. D. Song, S, Stobbe, P. Lodahl, \emph{Deterministic photon–emitter coupling in chiral photonic circuits}, Nature Nanotechnology volume 10, pages 775 - 778 (2015).


\bibitem{Lindell} I. V. Lindell, A. H. Sihvola, S. A. Tretyakov, A. J. Viitanen, \emph{Electromagnetic waves in chiral and bi-isotropic media} Artech House, Norwood, MA, 1994.


\bibitem{Cherenkov}F. R. Prudencio, M. G. Silveirinha \emph{Asymmetric Cherenkov emission in a topological plasmonic waveguide}, Phys. Rev. B 98, 115136 (2018).


\bibitem{Silv1sm} M. G. Silveirinha, \emph{Chern invariants for continuous media}, Phys. Rev. B, {92}, 125153, (2015).

\bibitem{Silv2} M. G. Silveirinha, \emph{Bulk-edge correspondence for topological photonic continua}, Phys. Rev. B, {94}, 205105, (2016).

\bibitem{Si_Re} S. A. Hassani Gangaraj, A. Nemilentsau, G. W. Hanson, \emph{The effects of three-dimensional defects on one-way surface plasmon propagation for photonic topological insulators comprised of continuum media}, Scientific Reports 6, 30055 (2016).

\bibitem{AliMulti} S. A. Hassani Gangaraj, M. G. Silveirinha, G. W. Hanson, \emph{Berry phase, Berry Connection, and Chern number for a continuum bianisotropic material from a classical electromagnetics perspective}, IEEE J. Multiscale and Multiphys. Comput. Techn., {2}, 3-17, (2017).

\bibitem{HeatTransportsm} M. G. Silveirinha, \emph{Topological angular momentum and radiative heat transport in closed orbits}, Phys. Rev. B, {95}, 115103, (2017).

\bibitem{HM_PRL} S. Ali Hassani Gangaraj and F. Monticone, \emph{Topological waveguiding near an exceptional point: defect-immune, slow-light, and loss-immune propagation}, Phys. Rev. Lett. 121, 093901, (2018).

\bibitem{HM_AWPL} S. A. Hassani Gangaraj, F. Monticone, \emph{Coupled topological surface modes in gyrotropic structures: green's function analysis}, IEEE Antennas and Wireless Propagation Letters,  DOI: 10.1109/LAWP.2018.2859796, (2018).


\bibitem{JPCM} S. A. Hassani Gangaraj, F. Monticone, \emph{Topologically-protected one-way leaky waves in nonreciprocal plasmonic structures}, J. Phys.: Condens. Matter 30 (2018).


\bibitem{Bittencourt} J. A. Bittencourt, \emph{Fundamentals of Plasma Physics}, 3rd ed. New York: Springer-Verlag, (2010).


\bibitem{nonlocal}  S. Raza, S. I. Bozhevolnyi, M. Wubs, and N. A. Mortensen, \emph{Nonlocal optical response in metallic nanostructures}, Journal of Physics: Condensed Matter {\bf 27}, 183204 (2015).

\bibitem{Shanhui_arxiv} S. Buddhiraju, Y. Shi, A. Song, C. Wojcik, M. Minkov, I. Williamson, A. Dutt, S. Fan, \emph{Absence of unidirectionally propagating surface plasmon-polaritons in nonreciprocal plasmonics}, arXiv:1809.05100 (2018).

\bibitem{Ishimaru} A. Ishimaru, \emph{Unidirectional waves in anisotropic media
and the resolution of the thermodynamic paradox}, Tech.
Rep. (Air Force Cambridge Research Laboratories, Bedford, Mass., 1962).

\bibitem{Mann} S. A. Mann, D. L. Sounas, and A. Alù, \emph{Nonreciprocal cavities and the time–bandwidth limit,}” Optica 6, 104 (2019).

\bibitem{note2} The specific value of the cover permittivity is not critical to observe the described effects: as long as the real part of the permittivity is negative the surface supports elliptic-like SPPs with no radiation leakage. We have verified that our results in Figs. \ref{EFS_1} would not qualitatively change if a different opaque material was considered (including a perfect electric conductor).

\bibitem{PRA_force} M. G. Silveirinha, S. Ali Hassani Gangaraj, George W. Hanson, Mauro Antezza, \emph{Fluctuation-induced forces on an atom near a photonic topological material}, Phys. Rev. A {\bf 97}, 022509,
(2018).

\bibitem{PRB_force} S. A. H. Gangaraj, G. W. Hanson, M. Antezza, M. G.
Silveirinha, \emph{Spontaneous lateral atomic recoil force close to a photonic topological material}, Phys. Rev. B 97, 201108(R) (2018).

\bibitem{PRB_torque} S. A. Hassani Gangaraj, M. G. Silveirinha, G. W. Hanson, M. Antezza, F. Monticone, \emph{Optical torque on a two-level system near a strongly nonreciprocal medium}, Phys. Rev. B 98, 125146 (2018).

\bibitem{note} We have also verified that, while it may be possible to obtain a similar response with a linearly-polarized dipole lying on the $xy$-plane and aligned with one of the beams, the intensity of the excited SPP beam is typically much lower, as the linear-dipole radiation does not match optimally the transverse angular momentum of the SPP beam.

\bibitem{CST} CST Microwave Studio 2018 (http://www.cst.com).


\end{thebibliography}
\end{document}